\documentclass[12pt,preprint]{aastex}
\def\lax {\ifmmode{_<\atop^{\sim}}\else{${_<\atop^{\sim}}$}\fi}  
\def\gax {\ifmmode{_>\atop^{\sim}}\else{${_>\atop^{\sim}}$}\fi}  
\def\gtorder{\mathrel{\raise.3ex\hbox{$>$}\mkern-14mu
             \lower0.6ex\hbox{$\sim$}}}
\def\etal { et al. }

\begin{document}

\title{On  the Nature of the Compact Object in SS~433.   Observational Evidence of   X-ray Photon Index Saturation}

\author{Elena Seifina\altaffilmark{1} and Lev Titarchuk\altaffilmark{2} }
\altaffiltext{1}{Moscow State University/Sternberg Astronomical Institute, Universitetsky 
Prospect 13, Moscow, 119992, Russia; seif@sai.msu.ru}
\altaffiltext{2}{Dipartimento di Fisica, Universit\`a di Ferrara, Via Saragat 1, I-44100 Ferrara, Italy, email:titarchuk@fe.infn.it; George Mason University Fairfax, VA 22030;   
Goddard Space Flight Center, NASA,  code 663, Greenbelt  
MD 2077, USA; email:lev@milkyway.gsfc.nasa.gov, USA}


\begin{abstract}
We present an analysis of the X-ray spectral properties observed  
from black hole candidate (BHC) binary SS~433. We have analyzed {\it Rossi} X-ray Time Explorer 
({\it RXTE})  data from this  source,
coordinated with Green Bank Interferometer/RATAN-600.
We show that SS~433 undergoes a  X-ray spectral transition from the low hard state (LHS)  to 
the intermediate state (IS).
  We show that the X-ray broad-band energy spectra during all spectral states are well fit by  
a sum of so called  ``Bulk Motion Comptonization (BMC)  component''  and
by   two (broad and narrow) Gaussians  for the  continuum and   line emissions respectively.
In addition to these spectral model components 
we also find a strong feature that we identify as a  "blackbody-like (BB)"  component which color 
temperature is  in the range of 4-5 keV in 24  IS spectra 
during the radio outburst decay in SS~433. 
 Our observational  results on the "high temperature
BB" bump  leads  us to suggest   the  presence
of  gravitationally redshifted annihilation line emission in this source. In fact this spectral feature   has been recently reproduced in Monte Carlo simulations by Laurent and Titarchuk.
We have also established the photon index saturation at  about 2.3 in  index vs mass accretion correlation.  This index-mass accretion correlation  
allows  us to evaluate the low limit of black hole (BH)  mass of compact object in SS~433, 
$M_{bh}\gax 2$ solar masses,  using  
the scaling method using BHC GX 339-4 as a  reference source.
Our estimate of the BH mass in SS 433 is consistent with   recent BH mass measurement using the radial-velocity measurements of the  binary  system by 
Hillwig \& Gies who find that  $M_{x}=(4.3\pm0.8)$  solar masses. This  is the smallest BH mass 
 found  up to now among all BH sources. 
Moreover, the  index saturation effect versus mass accretion rate revealed in SS~433, like in a
 number of other BH candidates,  is the strong observational evidence    
for the presence of a BH in SS~433. 
\end{abstract}


\section{Introduction}
The famous object SS~433 (V~1343 Aql) holds a special place in late 
twenty century Astronomy as the first microquasar discovered in  our Galaxy 
[see   reviews by \cite{margon84} and  \cite{fabr04}].
Observations of SS~433 have been carried out in all energy ranges  during more  than 
30 years. 
 Its key observational feature is  the 162.5 day precession period of the  jets 
that is revealed by the  lines features.
The radial velocity curves of these  lines 
are well described by a {\it Kinematical Model} 
which  reveals key parameters of the jets v=0.26c, and i=79$^\circ$ \citep{margon84}. 
Moreover, \cite{rom87} used and combined  these results with the  radio observations of the 
associated SNR W50 which  allowed  them to estimate distance of 5 kpc to SS 433.    

 SS~433 is an X-ray/optical 
binary with Algol type orbital eclipses.
This source is characterized  by two  distinct spectral states:  the quiescent hard state 
in which the persistent  jet flow takes place  and the soft  state when massive jet 
blobs are ejected~\citep{fielder87}. While the  quiescent  state has been 
well studied with numerous X-ray missions, 
only  
a few massive jet ejection  
events  were seen so far   simultaneously with  the X-ray soft  state 
[\cite{safiharb03}, \cite{band89}]. Because the ejection 
of a massive jet blob is  a rare (in average two times per year) and a short (approximately for ten days) 
event, we only  have observations for  the part of  this jet ejection.

SS~433 shows many kinds of variability related to regular 
(orbital and precessional) and irregular (flaring) activities. 
Although precession, binary orbital  and nutation 
 periods ($162^d$, $13^d.08$, $6.^d 28$ respectively)   are well known,  but a noticeable variability associated with 
shorter scales is  poorly investigated. It is worth noting that the  fast variability on time scale of  a few minutes 
was investigated by \cite{zwit91}  and \cite{gor87}
in the  optical V-band. 
More recently X-ray fast variability of 50 seconds time scale during flaring stage was 
found  using {\it RXTE} observations  by \cite{kot02}.
Power spectrum of SS~433 is well presented and approximated by power law 
 ($P\propto \nu^{-\alpha}$) in the range of the  $10^{-7}-10^{-2}$ Hz  according to X-ray timing data analysis  performed by \cite{rev06}.   They  also demonstrated  that   at frequencies lower than $10^{-5}$ Hz  the same variability pattern  takes place in  the
optical, radio and X-ray spectral bands (see Fig. 1 of that paper).

Many questions  regarding the complex behavior of SS~433 during outburst states as well as the nature of this compact  object and its
mass 
are still not  answered. However there is no shortage of  models  which are  
based on  radio, optical and X-ray variations of radiation detected from  SS~433 
 [see e.g. \cite{marsh02}, \cite{fabr04}, \cite{safiharb03}]. 



 The variation in mass estimates   of the compact ($M_x$),  optical ($M_v$) objects and 
the mass ratio ($q=M_x/M_v$) are quite broad.
\cite{kawai89} and later \cite{antokhina92} using   GINGA observations of  SS~433  estimated 
$q\simeq0.15$ and  $q=0.15-0.25$ respectively.
On the other hand \cite{kotani96} using  ASCA observations  
 found
$q\simeq 0.06-0.31$ in the frame of the precessing  jet model  with 
taking into account thermal adiabatic cooling of the jets  \citep{brinkmann91}.
Later   high-resolution 
observations by  \cite{gies02a}
found  the presence of absorption lines in 
the spectrum of the optical A ($\sim$A7Ib) supergiant companion. These  
orbital Doppler shifted absorption lines 
and stationary He~II emission from the companion allowed to estimate the mass ratio 
$q=0.35$,  implying the binary masses $M_x=4.3\pm0.8 M_{\odot}$ 
and $M_v=12.3\pm3.3 M_{\odot}$ in SS~433  [see \cite{hillwig08} for details]. 
Thus the average mass ratio inferred from this X-ray data analysis $q\le0.25$ 
is smaller than that inferred from optical observations $q\sim0.35$. 

One  comes to the conclusion that in the literature  there is a large variation in the mass estimates of the compact object ($M_x$), secondary star ($M_v$) and their mass ratio ($q = M_x/M_v$) in SS 433.
The nature of the compact  object  was inferred using the mass estimate or   its upper limit. 
No other strong arguments were used to determine the nature of the compact object in SS 433
which is an eclipsing X-ray binary system, with the primary most likely a black hole, or possibly a neutron star [see  e.g. \cite{cherep02}].   

In this work we 
 apply a substantially new approach for diagnosing  
the nature of compact object in SS~433.
In \S 2.1 we present details of radio  and X-ray observations of SS~433. 
Analysis of X-ray spectra are shown in \S 2.2.  We discuss  X-ray spectral evolution in SS 433  
in \S \ref{transitions}.  X-ray spectral properties  as  a function of orbital phase  are investigated  
in \S \ref{eclipses}. The results of  timing and power spectrum analysis are presented in 
\S \ref{timing}. We consider an interpretation of observational results   and show our arguments for BH 
presence  in SS 433  in \S \ref{theory}. We make discussion and concluding remarks in \S \ref{summary}.

\section{Observations and Data Reduction \label{data}}

\subsection{Listing  of X-ray observations used for data analysis} 
We analyzed the archival  data collected by PCA/{\it RXTE} \citep{bradt93} which were 
obtained in  time period from April 1996 to December 2006. These data allow us to investigate 
SS~433 in the broad X-ray 
energy band (3 -- 150 keV) during quiescent and outburst states. The {\it RXTE} data for SS~433
are available through the HEASARC public archive (http://heasarc.gsfc.nasa.gov) at the NASA Goddard
Space Flight Center (GSFC).  
As we
have already mentioned SS 433 shows continuous (associated with quiet state) and sporadic
(associated with active state) variability.
For investigation of the outburst state and  for comparing it with  the quiescent state  
we selected only observations during uneclipsed intervals of the binary orbital period.
In fact,  X-ray eclipse occurs around optical primary minima at phases $|\varphi|\leq0.1$.
 As a result we only used observations taken at 
interval $|\varphi|>0.1$ to exclude the eclipse orbital modulation.
 In total, this  type of observations includes 90 episodes  of outside  of eclipse phases. 
 Moreover, 27 observations 
 during eclipses taken at different precessional and orbital phases were used for spectral 
and timing analysis of orbital modulation effects.

Precessional ephemerids were taken from Fabrika et al. (2004). The moment of 
maximal separation between emission lines (T3)  was taken to be 
$T_3$=2443507.47 JD, the precessional period $P_{prec}$=162.375 days, 
the orbital period $P_{orb}$=13.08211 days, the moment of primary optical 
eclipse $T_0$=2450023.62 JD \citep{gor98}.

Standard tasks of the HEASOFT/FTOOLS
5.3 software package were utilized for data processing. 
We used methods recommended by  {\it RXTE} Guest Observer Facility  according to the {\it RXTE} Cookbook  (see http://heasarc.gsfc.nasa.gov/docs/xte/recipes/cook$~$book.html). 
For spectral analysis we used PCA {\it Standard 2} mode data, collected 
in the 3 -- 20~keV energy range. The standard dead time correction procedure 
has been applied to the data. To construct broad-band spectra,
HEXTE data 
have been also  used.
We  subtracted background corrected  in  off-source observations. 
To exclude the channels with largest uncertainties
 only data  in  20 -- 150~keV energy range were 
used for the spectral analysis. 
In Table 1 we list    the groups
 of {\it RXTE} observations covering  
 the source evolution  from  quiescent to outburst states. 
We also used public  data from the  All-Sky Monitor (ASM) 
on-board \textit{RXTE}. 
The ASM light curves (2-12 keV energy range ) were 
retrieved from the public \textit{RXTE}/ASM archive at HEASARC 
\footnote{http://xte.mit.edu/ASM\_lc.html}.




In the present Paper, we  have 
analyzed  X-ray spectra during quiescent and outburst states with reference to 
simultaneous radio and optical observations.  
The monitoring {\it RATAN-600 Radio Telescope} (2-8~GHz) data in the 1996 -- 2006 period were 
available through the public archive ({http://cats.sao.ru/$\sim$satr/BH}). 
We also  used radio observations by {\it Green Bank Interferometer, NRAO}\footnote{{http://www.gb.nrao.edu/fdocs/gbi/arcgbi}}  obtained from 1996 to 1998 at 2.25 and 8.3 GHz
and   simultaneous V-band photoelectric photometric  observations. Details of 
optical telescopes, reduction techniques and compilation methods are given by \cite{gor98}.


Additionally, we analyzed the INTEGRAL/IBIS/ISGRI spectra in flaring state (2004) of SS~433, that were
coordinated with the {\it RXTE} observations.  We have used the version 8.0 of the Offline Science 
Analysis (OSA) software distributed by INTEGRAL Science Data Center [ISDC, {http://isdc.unige.ch}, \cite{corv03}].

We  also present a comparison of the SS~433 data    with that for GRS~1915+105  obtained  during 
 {\it BeppoSAX} observations. 
We used two  {\it BeppoSAX}  detectors ({\it Medium-Energy Concentrator Spectrometer} (MECS) and a {\it Phoswich Detection System} (PDS)) for this analysis. 
The SAXDAS data package was utilized for performing data analysis. 
We  process the spectral analysis in the good response  energy range  
taking into account satisfactory statistics of the source: 1.8 -- 10 kev  for MECS 
and 15 -- 150 keV  for PDS.

\subsection{Spectral analysis}

SS 433 has long been of great interests in X-ray Astrophysics,
and has been observed early on with many satellites 
such as  HEAO-1 (Marshall et al. 1979),  EXOSAT (Watson et al. 1986), Tenma (Matsuoka et al.
1986), and Ginga (Kawai et al. 1989). 
Using HEAO-1, Marshall et al.  were the first to demonstrate that SS 433 is an
X-ray source. The HEAO-1 continuum was sufficiently modeled as thermal bremsstrahlung
with kT = 14.3 keV, and emission due to Fe-K was detected near 7 keV. 
The ASCA satellite, which carried X-ray CCD cameras for the first time, detected
many pairs of Doppler-shifted emission lines from ionized
metals, such as Si, S, Ar, Ca, Fe, and Ni, originating from the
twin jets (Kotani et al. 1994). 
The emission lines were also resolved with the Chandra HETGS, which
were found to have Doppler widths of  1000$-$5000 km s$^{-1}$
(Marshall et al. 2002; Namiki et al. 2003; Lopez et al. 2006).
The broad band continuum (up to 100 keV) is approximated
by a thermal bremsstrahlung spectrum with a temperature of
 10$-$30 keV, depending on whether SS 433 is in  or
out of eclipse (Kawai et al. 1989; Cherepashchuk et al. 2005).
Additional complex features were
detected from the XMM spectra, however, which could be
Compton-scattered emission from the jet base (Brinkmann
et al. 2005) or an iron-K absorption edge due to partial covering
(Kubota et al. 2007). From the width of an eclipse in
 the 25$-$50 keV band with INTEGRAL, Cherepashchuk et al. (2007)
and Krivosheyev et al. (2009) propose that a hot extended
corona around the accretion disk is responsible for the hard
X-ray emission via thermal Comptonization with a temperature
of  20 keV. High-quality X-ray spectra covering the broad
band are critical in establishing an interpretation of the high energy
spectra of SS 433.



\begin{figure}[ptbptbptb]
\includegraphics[scale=0.8,angle=0]{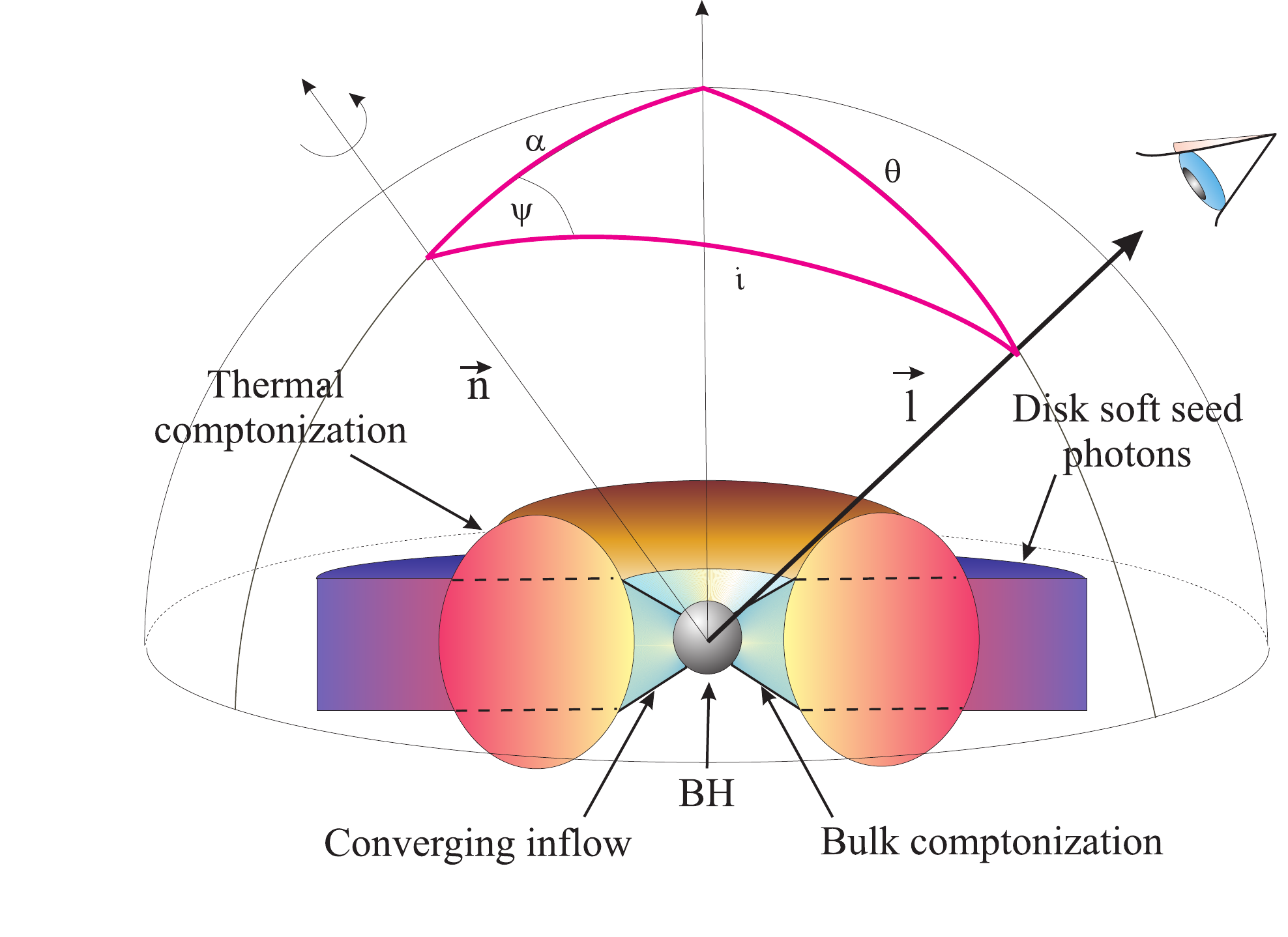}
\caption{
Schematic view of the proposed geometry for thermal and bulk Comptonization regions in the 
source SS~433 hosting a BH with PL-like emission at high energies. The terminal direction of viewing of 
the disk central part to Earth observer marks by vector $\vec l$. Here vector $\vec n$ is denoted 
normal to orbital plane, $i$ is  an inclination angle, $\psi$ is  a precession phase angle, $\alpha$ is 
cone precession angle and a bended arrow points the direction of the disc precession motion around 
normal $\vec n$ to the orbital plane. 
}
\label{geometry}
\end{figure}

In our study we model the broad band source spectra in  {\tt XSPEC} using an additive model consisting of  sum of the so called bulk motion Comptonization ({\it BMC}) model and two {\it Gaussian} line components.  The BMC model is a generic Comptonization model which can be applied to upscattering of soft photons injected  in a hot cloud. This model consists of two parts  where the first part is a direct blackbody (BB) component and the second one is a convolution of the fraction of the BB component  with a broken power law, the upscattering Green function.  
The spectral index of the blue wing  $\alpha$ is much smaller than that of the red wing 
$\alpha+3$.  The shape of the Green function (broken power law)  is generic and independent of the type of Comptonization, thermal or nonthermal.  The name of the model (BMC) has only a historical sense dating to 1997 [see \cite{tmk97}] when the model was first applied to case of  the bulk motion Comptonization. However this model can be applied to any type of Comptonization, thermal or nonthermal but it should be, in principle, combined  with exponential cutoff which is related to average plasma energy, 
for example, plasma temperature for the thermal Comptonization $kT_e$ or  kinetic energy of the matter in the case of the converging (bulk inflow) Comptonization.
In this Paper we consider a scenario related to our model (see Fig.~\ref{geometry})
where the Compton cloud  along with converging flow  are located
in the innermost part of the source and  a Keplerian disk  extends
from  the Compton cloud (CC) to the optical companion.

As we point out  ASCA and Chandra  detected many lines of various elements in the soft X-ray band of spectrum of SS 433. 
Particularly,  iron lines, Fe XXV -- Fe XXVI dominate at energies 6.5$<$E$<$7 keV
and these lines  show a  double structure due to jet Doppler shifts. 
In  addition to iron line  emission one can see the line emission related to 
 hydrogen - and helium-like ions of Mg, Si, S, Ar, Ca and Ni which   display  a double structure also.
 These line signatures  indicate that the lines are formed  in  the  relativistic jet configuration. Along with 
these lines there is an appreciable emission feature at 
6.4 keV which is visible in the X-ray spectrum   of SS 433~[\cite{kotani96}; \cite{seif00}]. 
This line is not subjected  to Doppler shifting.  Thus we want to emphasize that using the forms of these   lines  we can see features   of ``moving'' and ``stationary'' structures of material   surrounding  SS~433.

However the identification and precise theoretical reproduction of the line composition with 
{\it RXTE} is problematic because of its low energy resolution. 
As a test trial we added one 
Gaussian component to fit the spectrum 
varying the width and normalization of the line and we found that  the width $\sigma$ of this Gaussian feature roughly  ranges from 0.3 to 1 keV. In quite a few cases the    spectral fits using one Gaussian component  provide very wide residuals extended from  6 to 9  keV. 
However,  after  adding a second narrow Gaussian component (in the 6 -- 9 keV  range) 
the fit quality has been  significantly improved.
The energies of the first and second  Gaussian components $E_{line1}$ and $E_{line2}$ are presented  in  Table  \ref{tab:fit_table}.  For the first Gaussian $E_{line1}$  changes from 6.5 to 6.9 keV 
while the range of the second Gaussian varies from 7.1 to 9 keV. 
In some cases we see a wide residual taking place around 20 keV   which can  be a signature of  
"high temperature bbody-like"  spectral component  of  temperature in the range of  4-5 keV. 
Thus  we use our  {\tt XSPEC} model as {\it{wabs*(bmc+Gaussian+Gaussian+bbody)}} for  fitting of SS~433 spectra.   In particularly we use a value of hydrogen column $N_H=1.2\times10^{23}$ cm$^2$ which was found by \cite{Fil06} in calculations of   XSPEC model {\it wabs}.

The  best-fit parameters of  the source spectrum are 
presented in Tables 3 -- 6. 
For the {\it BMC} model  the  parameters  are
  spectral index $\alpha$ (photon index $\Gamma=\alpha+1$), 
color temperature of the blackbody-like injected  photons  $kT$, 
 $\log(A)$ related to the Comptonized fraction $f$ [$f=A/(1+A)$] and 
normalization of the blackbody-like component {\bf $N_{bmc}$}.
We find that color temperature $kT$ is  about 1 keV 
for all available {\it RXTE} data  and thus  we fix a value of $kT$ at  1 keV. 
When the parameter $\log(A)\gg1$ we fix $\log(A)=2$  (see Table 3, 5), because the 
Comptonized fraction $f=A/(1+A)\to$1. The variations  of  $A$ do not
improve fit quality any more. 
A systematic error of 1\% has been applied to the analyzed X-ray spectra. we applied this 
systematic error to the analyzed {\it RXTE} spectra in accordance to  the current version of  RXTE Cookbook (see http://heasarc.gsfc.nasa.gov/docs/xte/recipes/cook~book.html) following to
 recommended methods by {\it RXTE} Guest Observer Facility.

In Figure  \ref{geometry} we present a suggested geometry of X-ray 
 source in SS~433 (see more explanation of geometry below).  
 
 Similar to the ordinary {\it bbody} {\tt XSPEC} model,
the normalization $N_{bmc}$ is a ratio of the source (disk) luminosity
to the square of the distance $D$
\begin{equation}
N_{bmc}=\biggl(\frac{L}{10^{39}\mathrm{erg/s}}\biggr)\biggl(\frac{10\,\mathrm{kpc}}{D}\biggr)^2.
\label{bmc_norm}
\end{equation}  

The adopted spectral model shows  a very good performance throughout
all of the data set used in our analysis. Namely, the value of reduced $\chi^2$-statistic
$\chi^2_{red}=\chi^2/N_{dof}$, where $N_{dof}$ is a number of degree of freedom
for a fit, is less or around 1.0 for most observations. For a small 
fraction (less than 2\%) the  spectra with high counting statistic
$\chi^2_{red}$ reaches 1.5. However, it never exceeds a rejection 
limit of 2.0.

\section{Evolution  of 
SS~433  X-ray spectra  during outbursts
\label{transitions}}

We have established  general tendencies of quiescent-outburst behavior of SS~433 based on 
spectral parameter evolution of X-ray emission  in energy range (3-150) keV  using {\it RXTE}/P-CA/HEXTE data. 
We have also found correlation between  X-ray emission of SS 433 and  radio patterns observed 
by  RATAN-600 in the range 1-11 GHz and by {\it Green Bank  Interferometer} at 2.25,  8.3 GHz.

\begin{figure}[ptbptbptb]
\includegraphics[scale=1.0,angle=0]{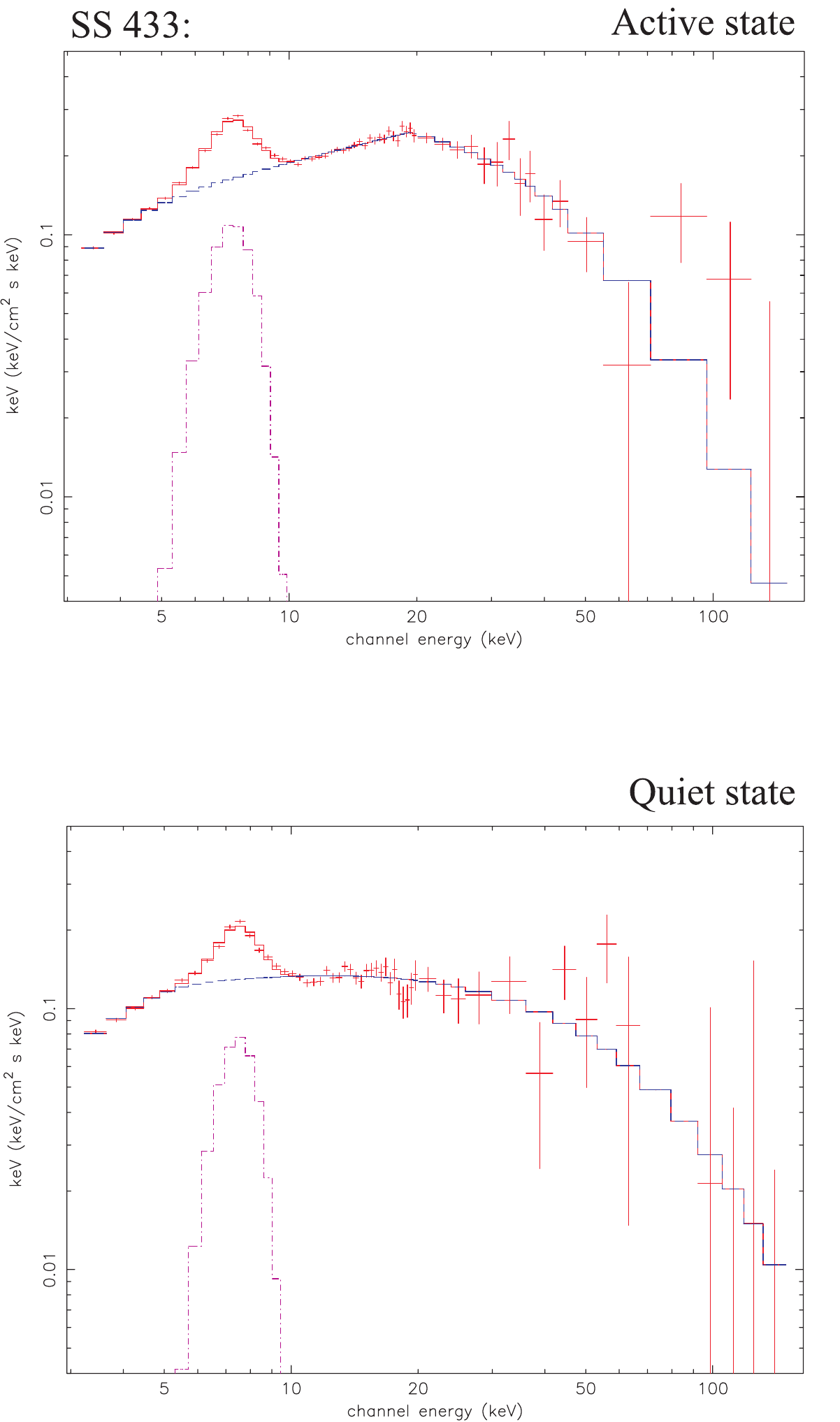}
\caption{ 
Two representative spectra of SS~433 in $EF(E)$ units  for 
 outburst IS ({\it top}) and for quiet IS ({\it bottom}). 
Blue line presents {\tt XSPEC} 
continuum component of the best-fit model (see text). 
}
\label{qui_act}
\end{figure}

We  identified  common  features  of ouburst behavior of SS~433 based on three available 
outbursts
and on two outburst decay 
sets. 
According to general BH state classification [see for example  \cite{ST09}, hereafter ST09]
SS~433  is mostly seen  in  the {\it intermediate} (IS) state.  
Different patterns of  SS~433 X-ray spectra
are listed at Table 3:
i) IS spectra  with $\Gamma\sim 1.9-2.2$ (see also  lower panel of   Fig.~\ref{qui_act}); 
ii) sum of BMC spectrum with $\Gamma\sim 2.2-2.3$ and  high temperature ``bbody''  component found  in 10-50 keV energy range 
(see Figs. \ref{2003_sp}-\ref{2005_sp}).

Ten days before radio  outburst, SS~433 being  in IS
 a stable  low soft (ASM) X-ray flux 
(see  Fig.~\ref{spec_evol_R2}) is  followed by 
a X-ray  flux rise reaching  its  maximum  just  two days before the radio flare.
Probably at the moment of radio peak,  MJD=50890  the X-ray flux reaches its minimum  
(BMC normalization, $N_{BMC}\sim0.7\times10^{-3}$)  and when X-ray spectrum becomes harder  (photon index $\Gamma\sim 1.9$).
It is worth  noting that spectral 
index  of radio emission $\alpha_{radio}$  has  its maximal value at a time   close to  X-ray outburst
(see  Fig.~\ref{spec_evol_R2}, at  MJD=50886).
Note the radio spectrum  is  harder  just a few days before 
 the  radio maximum. 
Then, after several days of  the radio maximum,   we see the X-ray   flux increases  while the 
radio flux substantially decreases, at least by factor 2 
 (see  Fig. ~\ref{spec_evol_R2} and Table 3).
 
 In Figs. \ref{spec_evol_R3}, \ref{spec_evol_R5} we  demonstrate   an evolution of the  radio,  X-ray flux and  X-ray spectral parameters   $N_{bmc}$ and  $\Gamma$  with time for different time intervals. 
 Notably in Fig. \ref{spec_evol_R3}   we   present the best-fit color temperature related to   ``high temperature BB''  bump which is  sometimes seen in IS spectra. 
Also  in Figs. \ref{spec_evol_R2}$-$\ref{spec_evol_R5} we show    a transition between intermediate and soft states given that    photon index varies  between 1.9  and 2.4. 

Thus  in Figures \ref{spec_evol_R2}$-$\ref{spec_evol_R5}  we present  the behavior of SS433 during 
  corresponding outbursts  and spectral transitions using  
i. the BMC normalization, ii. the photon index,  
iii.  ASM X-ray flux,  
and sometimes when they are available,  iv. GBI/RATAN-600 radio flux, 
v. radio spectral index (in case of GBI observations) and
  the color temperature (in the case of the occurrence of  `'high temperature Black body''-like feature).


The diagram of photon index $\Gamma$ versus BMC-normalization forms the track shown in 
Fig.~\ref{saturation} for all observations except   ones during orbital eclipses. The rising  part of 
the correlation  is seen  in both precession periods and outburst transitions. 
 The upper saturation part 
of the correlations  is  only seen during  outburst transitions. 
Note that in the cases of different outbursts of SS~433 
the levels of photon index saturation are also different. For example, the 2004 outburst 
(blue square) 
shows  the saturation level around $\Gamma\sim$ 2.3, while the  2005 -- 2006  outbursts
(crimson and red circles) saturates at  lower level of $\Gamma\sim$ 2.2. 
Comptonization  fraction $f$ shown in the right hand  panel of Fig.~\ref{saturation} is 
 high in most of cases. This means  that  in most  cases the soft disk  radiation  of SS~433 
 is subjected to reprocessing in a Compton Cloud  and only a small fraction of 
disk emission component ($1-f$) is directly seen.
Thus the energy spectrum of SS~433 during all states is dominated by a Comptonized component seen 
as a  power-law hard emission in the energy range from 3 to 70 keV, while 
the direct disk emission is not seen 
in all detected spectra.

\section{Spectral properties as  a function of orbital phase  \label{eclipses}}

\subsection{Detection of ``high temperature BB-like'' bump  \label{BB-bump}}

The ``BB-like'' feature has been  found  in 24 spectra of 
SS~433 among of all available data (R2, R3, R5-R7). 
Note that during  2004 outburst   (R5 set) ``high temperature BB-like''
 bump found  in  10-40 keV energy range of {\it RXTE}  has been  also detected  
 by ISGRI/IBIS detector onboard INTEGRAL satellite during simultaneous with {\it RXTE} observation
(ID 90401-01-01-01).  
For two time 
intervals (MJD=52225-52238, 53579-53588) this feature is detected in spectra 
near  the primary eclipse. It has been  visible before and after the  primary 
eclipse, while it is not seen during  the central phases of 
the primary eclipse. This screening effect with respect to orbital 
phases  is clearly seen in  two lower  panels (E3 and E4) of Fig.~\ref{hard_BB_eclipse}.
 Blue vertical strip marks  the interval of the  primary eclipse made according 
to optical ephemerids \citep{gor98}. Points marked with rose 
oreol correspond to spectra fitted by the model which includes "high-temperature BB-like" 
component (for details see Fig.~\ref{2003_sp}, \ref{2005_sp} and Tables 4-7). Red points in Fig. 
\ref{hard_BB_eclipse}
({\it see  two lower panels}) correspond to observations during primary eclipse 
when the  "BB-like"  component is not seen, while it has been found  before 
and after the eclipse.
Although observations are distributed 
more or less uniformly  with precession phase, the  "BB'' bump feature is seen 
generally at "disk face-on" position [$\psi=0.75$ (E3), $\psi=0$ (E4), 
Fig.~\ref{hard_BB_eclipse}], when the precessing 
disk is  open   and the 
innermost part of the disk is  more visible by the Earth observer. 

Thus we argue that  the "BB-like''  feature  is better detected at "disk face-on" 
position ($\psi\to 0$) than that  during   other precession phases. 
Moreover we have found  the total eclipse of "BB" bump during the  primary eclipse when the compact object is obscured by the normal star.
 In other words,  we  suggest that localization of "BB" bump 
feature emission should be better identified   during the direct observation of 
the  innermost 
  source  region near the compact object (CC). 

  However this ``BB'' component  can be smeared out by photon scattering if  
  optical depth of Compton cloud  is much greater than 1.  That  could be   the case during  observations
 for   MJD=50191-50194 (E1) and 50897-50907 (E2) (see Fig. \ref{hard_BB_eclipse}). 



\subsection{Softening of spectra at  some particular orbital phases}

The relative softening of the 
spectrum was found 
 in  the range of $\varphi=0.95-1.1$ for two orbital cycles of SS~433 (see Fig. \ref{hard_BB_eclipse}).
Softening of X-ray spectra     is seen when  the  compact hard emitting region, 
presumably related to  the central parts of the disc,   obscured  by the donor (optical) star and thus  
 the normalization of X-ray  spectrum  is suppressed due to  the eclipse.
This relative spectral softening is clearly seen 
at MJD=50191-50194 (E1), 50897-50907 (E2) 
(see two upper panels of Fig.~\ref{hard_BB_eclipse})
when  the noticeable  increase  
of photon index at MJD=50192.7 ($\Gamma=2.43\pm0.03$)  occurs with respect of that during  
the rest of  eclipse interval ($\Gamma=2.11\pm0.01$)  (see E1 panel in   Fig.~\ref{hard_BB_eclipse}). 
The similar behavior  of index is seen in  E2 panel where 
the noticable increase of photon index $\Gamma=2.39\pm0.01$ takes place at MJD=50899.7 with respect of that   when $\Gamma=2.13\pm0.01$ during  the rest of the eclipse interval. 

\section{Timing analysis  \label{timing}}

The {\it RXTE} light curves were analyzed using the {\it powspec} task from
FTOOLS 5.1. The timing analysis {\it RXTE}/PCA data was performed in the 2-15 keV energy 
range using  the {\it binned} mode.   We
generated power density spectra (PDS)
with 16-second time resolution. We subtracted the contribution due to
Poissonian statistics. To model PDS we used QPD/PLT plotting package.

We also analyzed  optical and radio light curves to compare them  with X-ray curve. 
Timing analysis of all available data of SS~433 demonstrates  power-law  PDSs of index around 1.5  
 in three energy bands (see Fig.~\ref{pow_3band}): X-ray ({\it blue}), 2.25 GHz radio ({\it black}), V-optical ({\it red}).  This result  is in an agreement with  previous results by \cite{rev06} who 
found that the SS~433 power spectra of  radio, optical and X-ray variabilities continue with the same power law from 
$10^{-7}$ Hz up to $10^{-5}$ Hz.

Such an extended 
behaviour of power spectra  is well known  for a number of X-ray 
binaries.
For instance,  \cite{churaz01}  found  that Cyg X-1 demonstrated 
the similar power law component at $10^{-5}-10$ Hz in X-ray energy band.  
An extensive study of low mass X-ray binary systems showed 
that such a power law behavior is common and   often  observed in LMXBs 
[see e.g. \cite{gilf05} and \cite{tsa07}]. 

The observed  variability  of SS~433 in 
different spectral bands was interpreted in the framework of self-similar 
accretion disk variations suggested by \cite{lyubar97}. 
In particular,  he  proposed that when 
variable mass accretion rate generates energy release in the disk far 
away from  a compact object  then the corresponding radiation  should be seen  in optical and UV   
energy range.    Whereas  the energy   emitted in X-ray 
 should be released near the compact  object.
In SS~433   presumably  these mass accretion rate fluctuations  are detected  and observed 
  in   power density spectrum.   In  which one can  also see that    
the X-ray variability correlates  with the optical  and radio emissions (see Fig. \ref{pow_3band}).

A study  of orbital modulation allow us  to detect some changes of power 
spectrum of SS~433  in the energy range 3 -- 15 keV during the primary eclipse.  
In the lower  panel of Fig.~\ref{spec_evol_R6}  we show  details of the typical evolution of 
X-ray timing and spectral characteristics during the primary eplipse. 
The top panels of this Figure  
 demonstrate 
 V-band optical light curve, BMC normalization and photon index $\Gamma$
for R6 set as a function of time (see Table 1 for details of these observations).   
In the plot of photon index $\Gamma$ versus time  points A and B correspond to 
moments MJD=53581 and 53585, during and after eclipse respectively.
Blue strips  mark eclipsed intervals of light curve. 
It is worth noting  that the optical and X-ray fluxes significantly drop   during the primary eclipse
 (see {\it two upper panels} and Table 3).

In Fig. \ref{spec_evol_R6} PDSs ({\it left bottom panel}) are plotted along with  energy spectrum
$E\times F(E)$ ({\it right bottom panel}) for two observational points A 
({\it 91103-01-03-00, blue}) and B ({\it 91103-01-07-00, red}).
The differences of the presented PDS and energy diagrams related to events A and B are noticeable.  
PDS during eclipse related to point A (blue histogram) indicates some weakening 
of power at 0.001 -- 0.01 Hz in comparison with  that for PDS outside of eclipse, at point B (red histogram). 
The  corresponding spectrum during eclipse (blue line)
demonstrates  a significant decrease of total flux  and also of the BMC normalization, at least by factor  2.  


\section{Interpretation of observational results \label{theory}}

Before to proceed with the interpretation of the observations let us to  briefly summarize our results  as follows:
i. The spectral data of SS~433 are well fit by a BMC model plus two Gaussian and ``high temperature bbody''  components (see \S 2.2).   
ii.  Power  law spectral index of BMC component rises and saturates 
with an  increase of the BMC normalization (disk flux). The photon index saturation 
levels of the 2004 and 2005 -- 2006 outbursts are about 2.2 and 2.3 respectively 
(see  Fig. \ref{saturation}).
iii. There is a total eclipse of "high temperature BB-like''  feature in the photon  spectrum because 
this spectral feature   is presumably formed  in  the innermost part of the X-ray source and thus  {\it it 
should not be  seen during the primary eclipse}.
 iv. The spectral index (hardness)   clearly changes    during the primary eclipse (see panels E1, E2 in Fig. \ref{hard_BB_eclipse} and one of upper panels in Fig. \ref{spec_evol_R6}).  
 This   can be   explained if there is  a  slight variation in spectral shape with radius  through  X-ray spectral formation region.
  
\subsection{Index saturation as a BH signature.  BH mass  in SS~433}
\label{index saturation}
We have firmly established that  the photon  index  of the resulting spectrum of SS~433 saturates with the BMC normalization $N_{bmc}\propto L_d$ 
or with disk mass accretion rate.  ST09 presented strong arguments that this index saturation is 
a signature  of converging flow into BH.  In fact, the spectral index $\alpha$ is inverse proportional to Comptonization parameter $Y$ which is proportional to  product of average number of photon scattering in Compton cloud $N_{sc}$ and mean efficiency of gaining energy at  any scattering $\eta$. 
Because the index saturates, 
it means $\eta$ is an inverse of $N_{sc}$.
But   in general for an optically thick Compton cloud of optical depth $\tau\gg1$  average number of scatterings  $N_{sc}$ is proportional to  $\tau^2$  or $\tau$ [see e.g. \cite{LT99}].   
 In the case of thermal Comptonization $N_{sc}\propto \tau^2$ and then $\eta$ should be inverse proportional to  $\tau^2$. But in the thermal Comptonization case   the efficiency $\eta$ is proportional to electron temperature $kT_e$ only [see e.g. \cite{st80}]   and it is not clear why  there could be such a sophisticated  tuning   in the Compton cloud that $kT_e$  steadily decreases with $\tau$  as 
 $\tau^{-2}$. 
On the other  hand $N_{sc} \propto \tau$ and $\eta\propto 1/\tau$ in the  converging flow case 
[see e.g. \cite{tmk97} and ST09] and thus the index saturation is a natural consequence of these laws for 
 $N_{sc}$  and $\eta$ in the case of converging flow [see \cite{tz98} for the solution of the full relativistic kinetic equation].     Furthermore ST09 demonstrate  that for a number of BHC sources index saturates not only when the mass accretion rate increases but also when QPO frequency increases. These index correlation and saturation with QPO frequency are a direct evidence that  the index saturation takes place when the size of Compton cloud (CC) becomes more compact (namely the index saturates when compactness of the emission area increases). 

  \cite{tlm98} demonstrated that the CC compactness  increases with mass accretion rate, 
  $\dot M$.  
   Moreover in the BH case the shape of the emergent spectrum saturates, when $\dot m$ increases, to the spectra for which the power-law indices $\Gamma$ stay  almost the same and these $\Gamma$ are always below 3 (see ST09 and Fig. \ref{saturation} in this Paper). In fact,  the asymptotic  values of $\Gamma$ in SS 433 vary  in the ranges $\sim 2.2-2.3$ 
(see Fig. \ref{saturation}). Thus, we  argue that the SS~433 X-ray observations 
reveal the index saturation vs mass accretion which can be a sign of a converging flow in this source
indicating a BH presence. 
This index vs  BMC normalization (proportional to disk mass accretion rate) correlation  can, in principle, allows us  to estimate a lower limit of  BH mass  in SS 433.
 
 



To do BH mass estimate in SS~433  
we choose to  scale index-normalization previously  found  in GX~339-4 (see ST09) with that in SS~433
(see Fig. \ref{gx339_scal}). We can proceed with this scaling if these two correlations  are self-similar. 
In other words when given correlations have the same the index saturation levels and slopes as a function of normalization which is proportional to  mass accretion rate in the disk. 

Then the value of the scaling coefficient provides us BH mass estimate. 
Note that the index-normalization correlation curve  for  2003 decay data 
of GX~339-4 (taken from ST09) is self-similar with that we find  for SS~433.


Use of scaling method for  these two correlations provides the scaling coefficient  $S_N=14.9\pm0.7$ which is a ratio (see ST09):
\begin{equation}
s_N=\frac{N_r}{N_t}=\frac{M_r d^2_t}{M_t d^2_r}f_G
\label{scaling coefficient}
\end{equation}
where $M_r$ is BH mass of the reference source (GX 339-4), $M_t$ is BH mass of the target source 
(SS 433), geometric factor $f_G=(\cos\theta)_r/(\cos\theta)_t$, inclination angles $\theta_r$,  $\theta_t$ and $d_r$, $d_t$ are distances to the reference and target sources. Here the subscripts $r$ and $t$ are related to the reference and target sources respectively. 
One can, in principle, estimate BH mass $M_t$ in SS 433 using the relation (\ref{scaling coefficient}).
However a value of geometric factor $f_G=(\cos\theta)_r/(\cos\theta)_t$  is not obvious because
$\cos\theta_r$  which is an inclination angle of X-ray emission area in GX 339-4 is not known. 
Also it is not clear if we can use the inclination of the whole disk in order to estimate the inclination of the innermost part of the disk which size is about 4 orders of magnitude less than that of the disk.
This is one of the main problems to correctly evaluate $f_G$. 

In order to estimate $(\cos\theta)_r=(\cos\theta)_{GX}$  we can apply the same  formula (\ref{scaling coefficient}) but using the  target and reference sources  GX 339-4 and GRO J1655-40 respectively. 
We take  values of $M_r$, $M_t$,  $d_r$,  $d_t$  and $\cos \theta_{GRO}$ from ST09 (Table 5 there)  
and then we obtain the lowest limit of 
\begin{equation}
(\cos\theta)_{GX}\gax 0.4\frac{(m_r/6.3) (d_t/ 5.75~{\rm kpc})^2[(\cos\theta)_{GRO}/\cos70^0]}{(m_t/12.3) (d_r/ 3.2~{\rm kpc})^2}, 
\label{scaling coefficientm}
\end{equation} 
Now we can resolve formula  (\ref{scaling coefficient})  with respect to $m_t=m_{ss}$ where we use  GX 399-4 and  SS 433  as the reference and target  sources respectively.  Thus we obtain 
\begin{equation}
m_{SS}\gax1.8\frac{(D_{t}/5.0~{\rm kpc})^2}{(D_{r}/5.75~{\rm kpc})^2} 
\frac{[(\cos\theta)_{GX}/0.4]}{[(\cos\theta)_{ss}/0.17]}
 \frac{(m_{r}/12.3)}{(s_N/14.9)}
\label{SS Mass}
\end{equation} 
where we use a value of $\cos\theta_{ss}\approx0.17$ estimated by \cite{cherep05}.

This is the lower limit  mass estimate for SS~433 found using the ST09 scaling method. 
 This estimate  is consistent with the BH mass value 
 $M_{SS}=(4.3\pm0.8) M_{\odot}$ obtained using optical observations of SS~433
[see \cite{hillwig08}, hereafter HG08].
BH mass  estimates  for the  reference (GX 339-4) and target (SS 433) sources   are summarized  in Table~\ref{tab:par_scal}. 





\subsection{
Upper limit of  the size of ``High temperature BB-like'' bump region  \label{structure}} 

Now we try to estimate   the upper limit of the size of the region $\Delta l_{BB}$  from  where the  " high temperature BB-like'' photons originates.  We consider the screening of the  "BB-like'' bump emitting region by 
the optical star with radius $R_{st}$.  
 
Based on BH mass estimate  $m_x$ by  HG08,  one can determine the 
mass of donor (optical) star using  the mass function of normal star 
$f_v(m)$ for SS 433. Namely,
\begin{equation}
f_v(m)=\frac{{m_x}^3\sin^3 i}{(m_x+m_v)^2}=1.038\times 10^{-7}{K_v}^3*P_{orb}(1-e^2)^{3/2},
\label{mass_func}
\end{equation}
where $K_v=58.2\pm3.1$  km s$^{-1}$ is  semi-amphlitude of radial velocity, $P_{orb}=13^d.08$ is  
orbital period in days,  
$e$ is  eccentricity  ($e\sim 0$), $i=78^\circ .8 $ is  inclination angle of system  [see \cite{margon84}]. 
Thus mass of the donor star $m_v$ can be obtained as
\begin{equation}
m_v=\sqrt{\frac{m_x^3 \sin^3 i}{f_v(m)}}-m_x\simeq12.3~M_{\odot} 
\label{donor_mass}
\end{equation}
 if one resolves Eq.(\ref{mass_func}) with respect to 
$m_v$  and uses the  BH  mass values found by HG08
(see above).
Then  the resulting mass ratio is $q=m_x/m_v\le 0.35$
 [see HG08 and \cite{Fil06}].

Using    the donor star mass one can calculate the binary separation 
and Roche lobe radius. The resulting binary separation from Kepler's 
third law is $a\sim60 R_{\odot}$. Because the donor (optical) star in SS~433 fills its Roche lobe 
 $R_{st}=R_{cr}$, the  Roche lobe radius of the  donor 
star can be found  (compare with Eggleton, 1983)
\begin{equation}
R_{st}\sim  a\times\frac{0.49 q^{-2/3}}{0.6 q^{-2/3}+\ln(1+q^{-1/3})}\sim28R_{\odot}.
\label{ Roche_lobe_radius}
\end{equation}

But the upper limit of  the "BB-like'' bump  forming region can be estimated as  
\begin{equation}
\Delta l_{BB}\sim \frac{2\pi a\Delta t_{BB}}{P_{orb}}-2*R_{st}<6.8 R_{\odot}.
\label{size_BB_region}
\end{equation}
using that  "BB-like'' bump forming region is screened,  during $\Delta t_{BB}<$ 2.5 days, 
by a  donor (optical) star of radius $R_{st}$ and rotating with   the orbital velocity $v=2\pi a/P_{orb}$ 
 (see Figs. \ref{hard_BB_eclipse}, \ref{eclipse_picture}). 
It is worth  noting  that $\Delta l_{BB}$,   the real size  of  ``BB-like'' component emission,  should be significantly smaller than that we estimate by inequality  (\ref{size_BB_region}). 
Here we have  just found the upper limit of $\Delta l_{BB}$   
based on the direct observational estimates.


\section{Discussion  and Summary \label{summary}} 

We derive a lower limit of 2 solar masses for the compact object in SS 4333 which clearly does not rule out a neutron star - this limit alone does not give strong observational evidence for a black hole. 
This is  the lower limit of the central  object   mass  is obtained using sliding the index-mass accretion correlation of SS 433 with respect to that for GX 339-4. 
In fact, this is an example that the sliding coefficient provides the lower limit of the mass of the central object.  We remind a reader that this correlation of photon index and mass accretion rate was obtained using the  {\it so called} BMC model of XSPEC which is the first principle radiative transfer model. Namely we consider and confirm applying the XSPEC fitting procedure that
the emergent spectrum of the Compton cloud (CC, the innermost part of a BH) is a result of upscattering of soft, presumably disk,  photons off  CC energetic electrons. 

The crucial characteristic of the correlation is the saturation plateau which is a signature of the converging flow into central object because the spectral index is an inverse of Comptonization parameter  $Y$ which is a product of number of scatterings  $N_{sc}$ and average fractional energy change per scattering $\eta$ (see also details  of explanation of index saturation in \S \ref{index saturation}). 
Only for converging flow $N_{sc}$ is proportional to optical depth $\tau$ of the cloud and $\eta$ is inverse proportional to $\tau$ when $\tau$ is greater than one. Thus  for large $\tau$ (mass accretion rate) it is natural to expect the index saturation as a signature of  a BH and  in fact, we really see this saturation  in the case of SS 433. In the other words the main evidence for black hole is the index saturation  with mass accretion rate. In contrast, in neutron star sources the photon index is almost constant, concentrating near  2, i.e, it   does not change  when mass accretion rate increases
[see \cite{ft10}]. 

Moreover \cite{hillwig08} 
found, using optical observations of SS 433,  that the mass of central object is $4.3\pm 0.8$  solar masses  which can be the case  only for a black hole. It is also worth noting a recent paper by 
\cite{kub10} who argue these mass values should be  considered  as upper limits and taking  into account  the heating of the donor star may lower the derived masses to 
$1.9M_\odot\leq M_X\leq  4.9 M_\odot $.

\begin{figure}[ptbptbptb]
\includegraphics[scale=1.,angle=0]{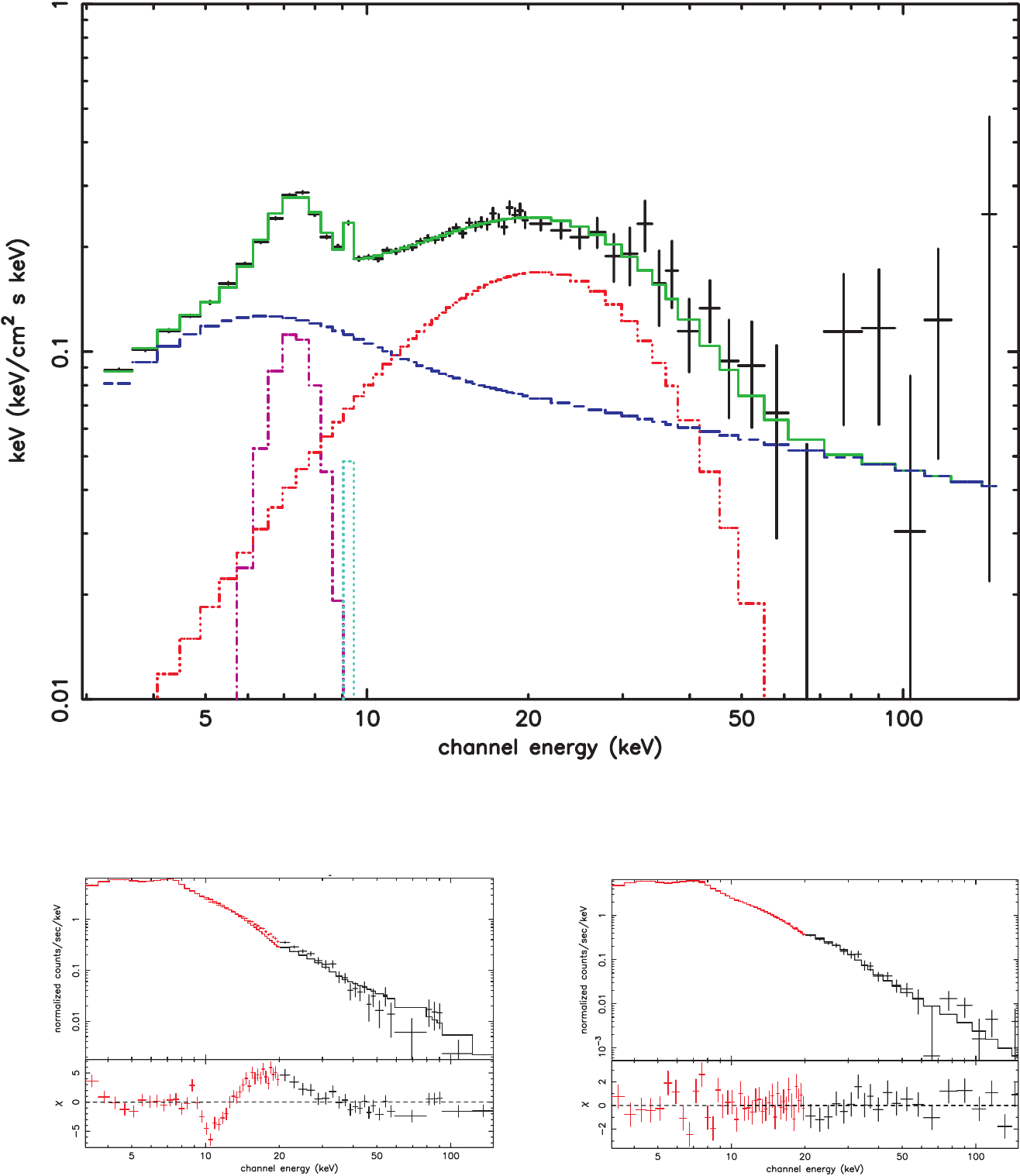}
\caption{
The best-fit spectrum during of 2004 outburst  transition
in $EF(E)$ units ({\it top}) and in normalized
counts units ({\it bottom panel}) with $\chi$ for the 90401-01-01-01
observation (R5 set). {\it Left}: the best fit which does not include   a high-temperature
``bbody'' component
($\chi^2_{red}$=4.05 for 72 d.o.f.) and {\it right}: the best-fit spectrum and
$\Delta\chi^2$, when
residual hump at 20 keV is modelled by a high-temperature ``bbody''
component with $T_{BB}=5.9\pm0.2$ keV  ($\chi^2_{red}$=1.00 for 70 d.o.f.). On top
panel data are denoted by black points. Model  consists of  four components  shown  by
 blue, 
 dashed purple,  
  light blue,  
 and red  
 lines  for {\it BMC}, {\it Gaussian1}, {\it Gaussian2} and {\it bbody} components
 respectively.
}
\label{2003_sp}
\end{figure}

\begin{figure}[ptbptbptb]
\includegraphics[scale=1.,angle=0]{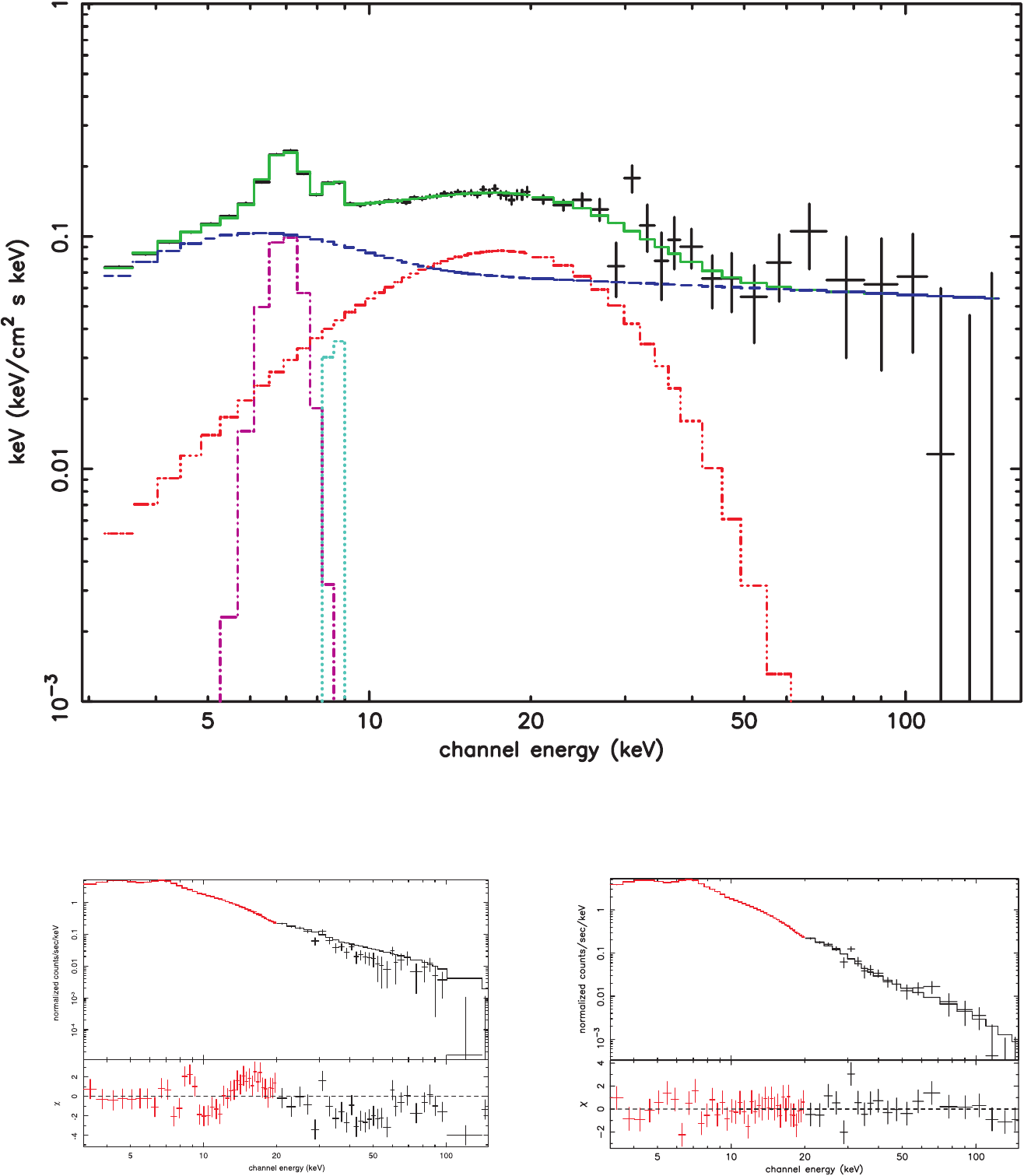}
\caption{
The best-fit spectrum during of 2005 outburst  transition
in $EF(E)$ units ({\it top}) and in normalized
counts units ({\it bottom panel}) with $\chi$ for the 91092-01-01-00
observation (R6). {\it Left}:  the best fit which does not include   a high-temperature
``bbody'' component
($\chi^2_{red}$=5.01 for 72 d.o.f.) and {\it right}: the best-fit spectrum and
$\Delta\chi^2$, when
residual hamp around 14 keV is modelled by a high-temperature bbody''
component  with $T_{BB}=4.5\pm 0.2$ keV ($\chi^2_{red}$=1.07 for 70 d.o.f.). On top
panel data are denoted by black points.
Model  consists of  four components  shown  by
 blue, 
 dashed purple,  
  light blue,  
 and red  
 lines  for {\it BMC}, {\it Gaussian1}, {\it Gaussian2} and {\it bbody} components
 respectively.
}
\label{2005_sp}
\end{figure}

\begin{figure}[ptbptbptb]
\includegraphics[scale=0.9,angle=0]{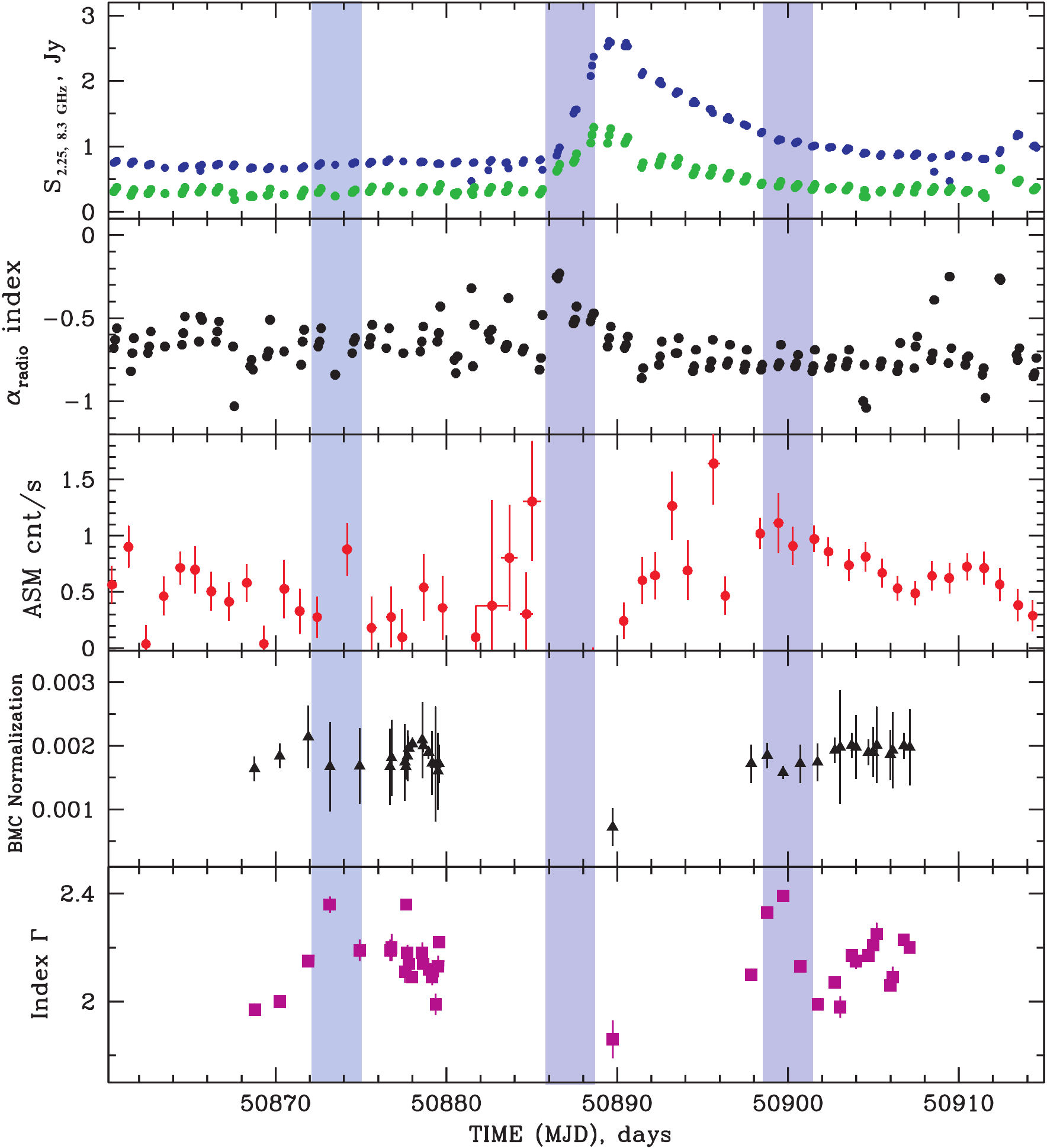}
\caption{
From top to bottom: Evolution of the flux density (GBI, 2.25 GHz -- blue, 
8.3 GHz -- green), 
RXTE/ASM count rate, 
spectral index $\alpha_{radio}$ in corresponding radio GBI range, 
BMC normalization and photon index $\Gamma$ for the middle of 
2004 outburst of SS~433. Blue strips mark eclipsed intervals of light curves 
(around MJD 
50872-50875, 50886-50889 and 50899-50902). 
}
\label{spec_evol_R2}
\end{figure}

\begin{figure}[ptbptbptb]
\includegraphics[scale=0.9,angle=0]{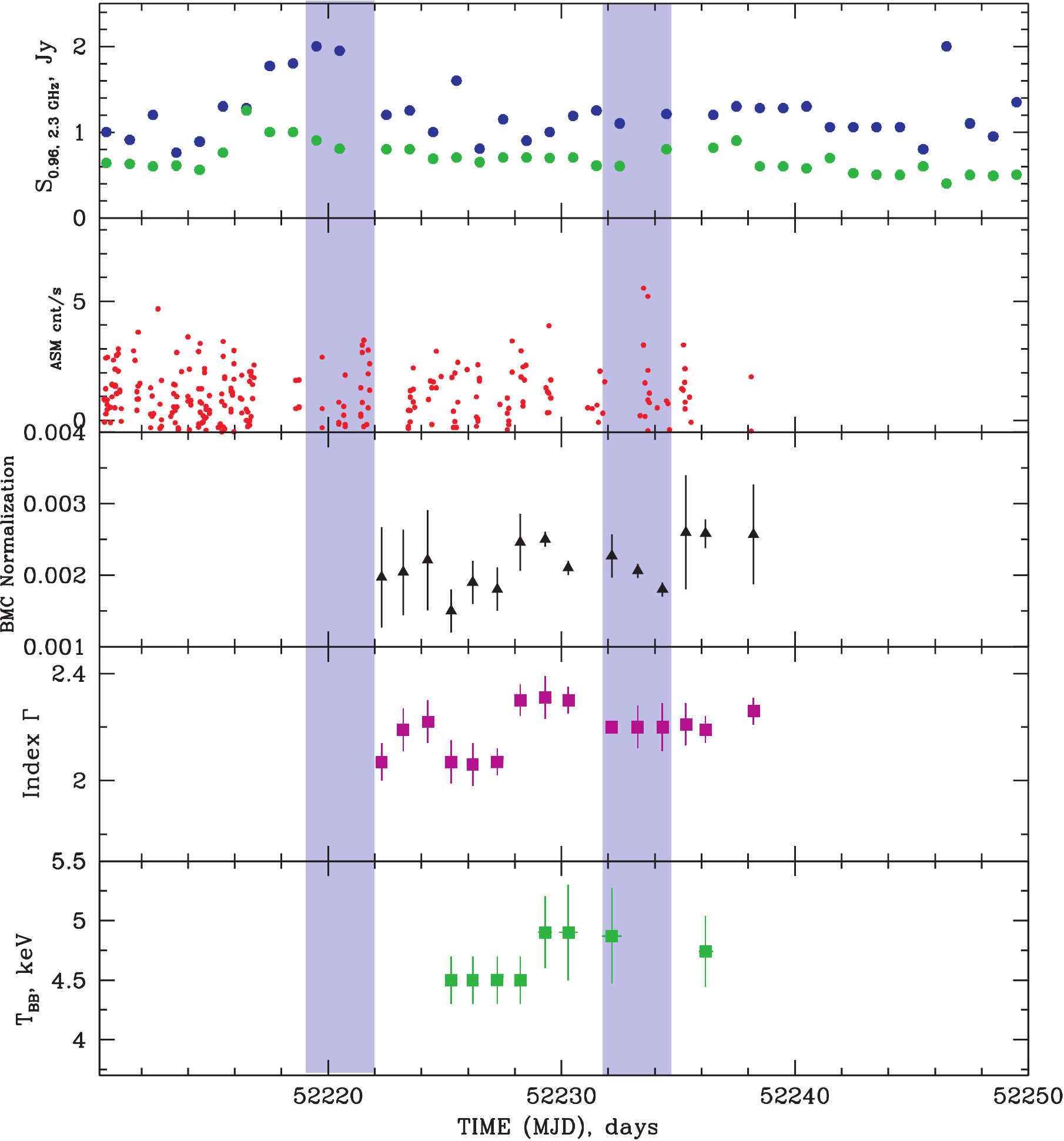}
\caption{
From top to bottom: Evolution of the flux density 
[0.96 ({\it green}) and 2.3 ({\it blue}) Gz),
 RATAN-600], RXTE/ASM count rate, BMC normalization, photon index $\Gamma$ and color temperature $T_{BB}$ of high temperature ``Black Body'' spectral component for the middle of 2001 outburst of SS~433. Blue strips mark eclipsed intervals of light curve around MJD 52219 -- 52221 and 522222 -- 52234 during R3 set.
}
\label{spec_evol_R3}
\end{figure}

\begin{figure}[ptbptbptb]
\includegraphics[scale=0.9,angle=0]{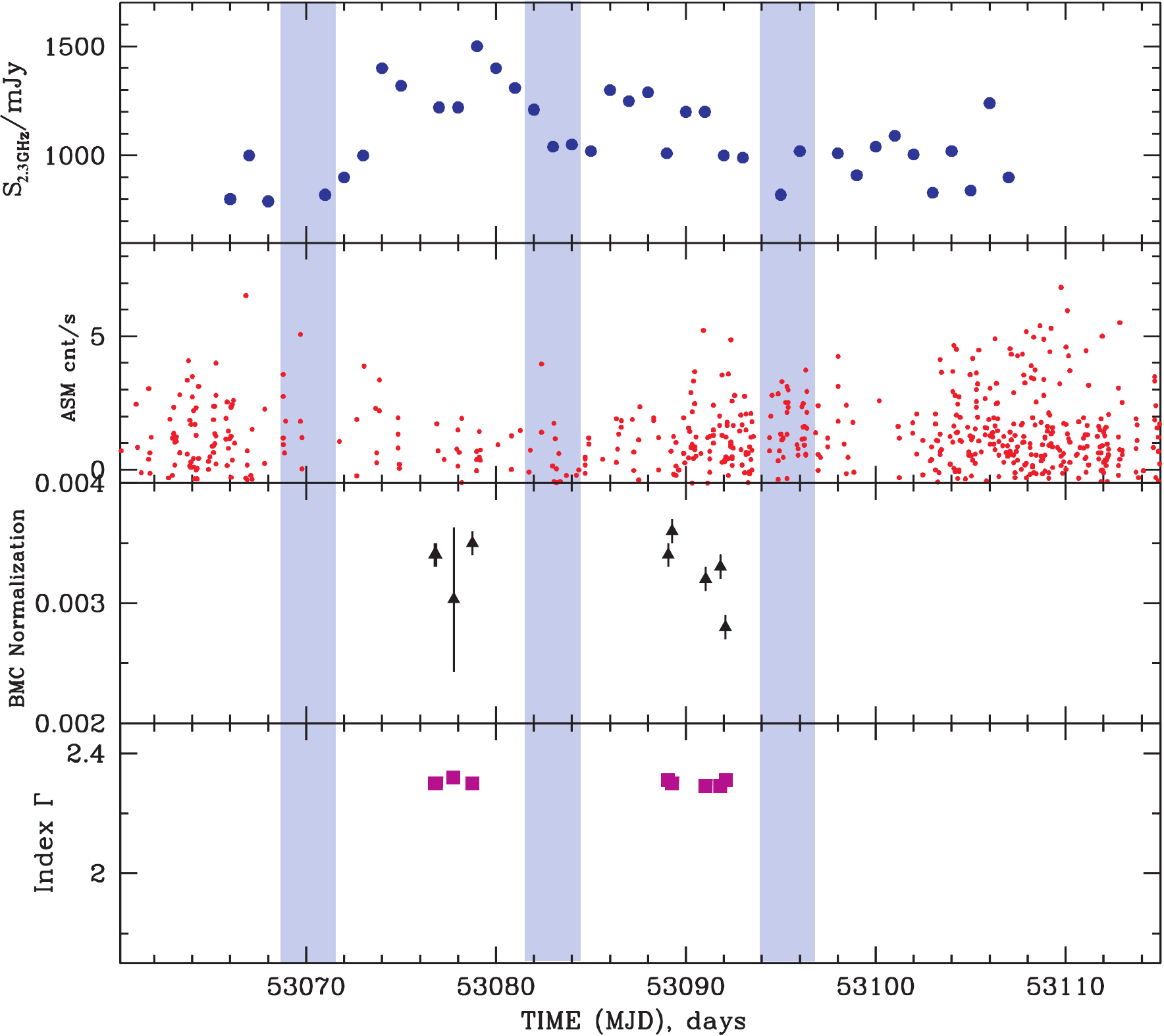}
\caption{
From top to bottom: Evolution of the flux density (2.3 GHz, RATAN-600), RXTE/ASM 
count rate, BMC normalization and photon index $\Gamma$ for the middle of 
2004 outburst of SS~433. Blue strips mark eclipsed intervals of light curve 
around MJD 53070 -- 53100 (R5 set). 
}
\label{spec_evol_R5}
\end{figure}

We find that the BMC model  along with an additional  ``high temperature bbody-like'' component allow  us to consistently   describe X-ray spectra of SS~433  and derive  physical conditions during all states of SS~433. Moreover, our approach for  data fitting  demonstrates 
 that  the SS~433 spectra  becomes  softer and finally photon index saturates   when mass accretion rate increases during  X-ray outburst. 
This index behavior is in agreement  with that ST09 and TS09 previously find in a number of other BHC sources. 
Furthermore,  the   fits,  using  our spectral model, allow us  to demonstrate that in some cases 
 an additional  ``high temperature bbody -like''  component  of color temperature of 4-5 keV  appears  in the data (see Figs. \ref{2003_sp}-\ref{spec_evol_R5} and Tables  4-6). 
 This  ``bbody-like'' feature 
arises only during outburst transition of SS~433 and it presumably  originates in   the  
innermost  part of accretion flow (see discussion in section \ref{structure}). 

Recently  we found a   signature  of this "high temperature bbody-like" bump in  {\it RXTE} spectral data 
 for  BHC  GRS 1915+105   (see TS09).
 Also  we  find  that the same kind of   ``blackbody-like''  bump      is seen in   BeppoSAX observations of the  intermediate state of  GRS 1915+105  (April 21, 2000, ID=209850011).
 The data along with the model  are presented in Figure~\ref{sp_sax_int} ({\it top panel}).
Consequently  the presence of the additional  ``BB-like''   bump spectral feature  is not an instrumental effect of  {\it RXTE} but it can be  a real observational feature of  the observed  spectra of BHCs.
Furthermore, we also detected the same `bbody-like 
high temperature'' feature in the  spectrum of SS~433 using IBIS/ISGRI detector 
onboard INTEGRAL satellite during simultaneous observations with {\it RXTE} 
(ID=90401-01-01-01, R5 set)  (see Fig.~\ref{sp_sax_int}  {\it bottom panel}).  
Thus it is quite possible that this ``high temperature bbody-like " feature can be the universal signature of  BH spectra. 

Also one can argue that the presence of ``high temperature bbody-like" feature may be an artifact of having chosen the BMC model. In fact, the presence or absence of the bump can be related to the model.  But  the BMC model is a first principle Comptonization model and therefore any residual using this model can be considered as an observational signature of other  component presented in the spectrum.  Moreover \cite{LT99} and \cite{lt10} demonstrate using  Monte Carlo simulation  that  X-ray emergent spectra of a black hole  for all spectral states   are  well fitted by BMC model   if the effect of pair production is not taken into account. But in the spectra of the  intermediate and soft states this  additional ``BBÕÕ features  appears if the non-linear  effect  of pair creations in very vicinity of a black hole horizon is taken into account [Laurent \& Titarchuk (2010) in preparation]. 

One can say that such a bump seems very similar to the ``Compton-bump'' expected by reflection models in black hole systems (both AGN and stellar-mass objects), e.g. see  \cite{mil07}.
In fact, in our previous paper TS09  (see section 4.2 there) we discuss this issue in detail. Here we should once again  point out  that in the observations this ``high temperature BB-likeÕÕ bump appears only when the photon index is higher than 2. On the other hand \cite{lt07} 
demonstrate using Monte Carlo simulation and theoretical arguments that the reflection bump never appears in the emergent spectra if  photon index of  the spectrum $\Gamma$ is higher than 2.  Namely, there are not   enough photons in  the incident spectrum at high energies, if $\Gamma>2$,    to be reemitted  into lower energies due to scattering and  recoil effects.   
In fact, as one can see from Tables 4-6, that in all spectra where we detect this 
``high temperature  BB-like'' ($\sim20$ keV)  feature index of the hard BMC component $\Gamma > 2$ and thus the appearance of this bump in the observed spectra cannot be explained by so called ``reflection'' effect.  

TS09 argue that this ``high temperature  BB-like'' bump can be a result of the gravitationally redshifted annihilation line which is {\it originally} formed in the very vicinity of a BH horizon. 
This feature (bump) should be seen in the IS and HSS spectra only because  in these states  mass accretion rate is high enough to provide conditions to form the strong annihilation line and also to observe this feature through accreting material of  relatively low plasma temperature (see more details in TS09, \S  4.2 there).  

It is worth noting that a time delay of the radio peak with respect  to  that in X-rays  is   two days for 
outburst rise \citep{rev06}.
We find  that during  this X-ray flare, and then in  radio one,  the object  transits 
from IS to LHS, which is  quite unusual  for spectral transitions of BHC. 
This particular  behavior of SS 433 is different from other BH sources probably because  the jet is the dominant emission component.
In general, when a BH goes to outburst it leaves a quiescent state,  enters to LHS  and then it goes to IS-HSS.  
However,   the  microquasar SS~433   is   the only   BHC  which stays,   most of the time  in  IS 
with a rare  short transition to LHS  accompanied by the radio flare. 
A delay of radio peak with respect  to 
X-ray peak is known  during outburst rise transition in  many X-ray BHC. 
However a time delay value is different for each of the sources. For instance,  it 
is about 20 days for GRS  1915+105 and  about 2 days  during outburst in  SS~433.
 If in the  GRS 1915+105 case the index saturation value is about 3, whereas  in the case  of SS~433 the index saturates  to $\Gamma\sim 2.3$.
Moreover  the direct soft component, which is  usually associated with the disk,
is clearly seen in  the GRS 1915  soft spectra and it   is weak or absent at all 
(at the level  of detection) in most of  SS~433 spectra (see  e.g. Table 3).



\begin{figure}[ptbptbptb]
\includegraphics[scale=1.,angle=0]{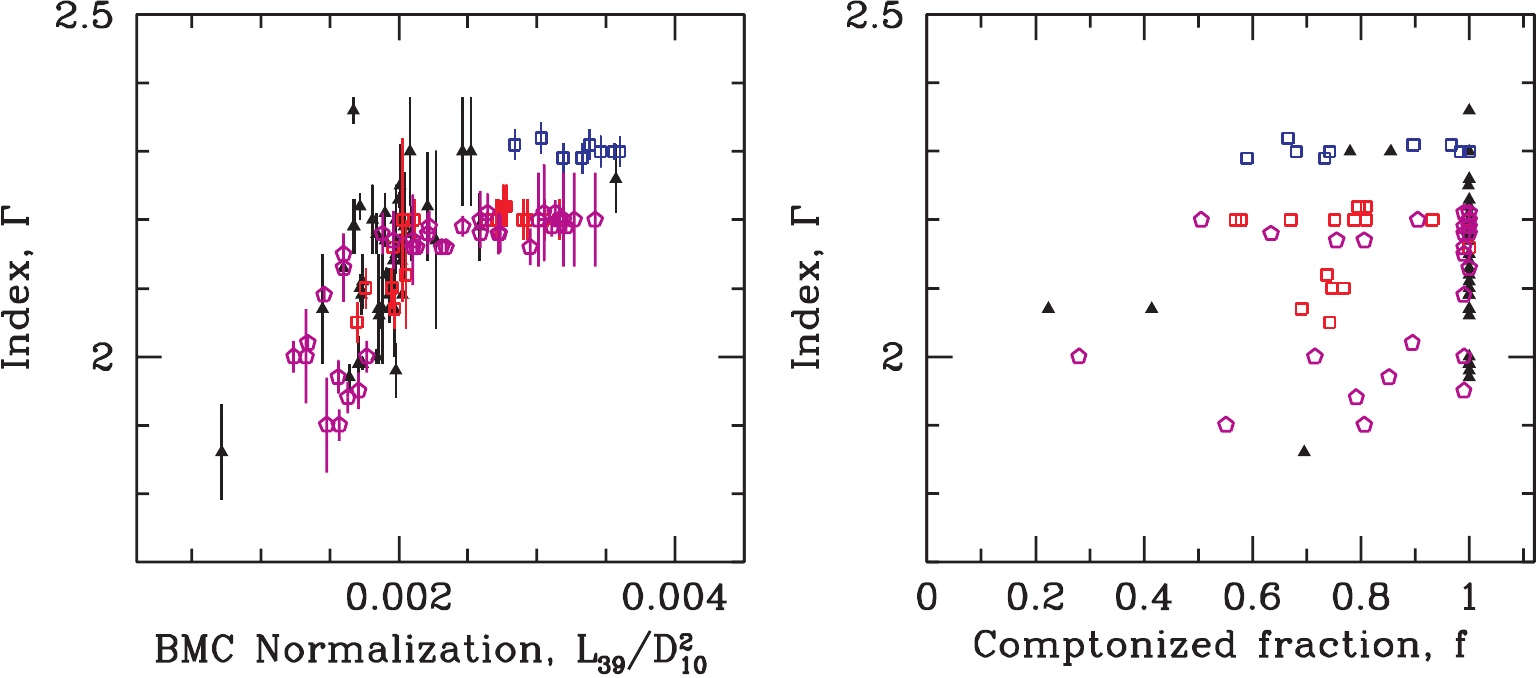}
\caption{
Photon index $\Gamma$ plotted versus BMC normalization ({\it left}) and 
Comptonized fraction ({\it right}) for all observations. Blue squars correspond to 
2004 outburst. Crimson circle marks 2005-- 2006 observations. Red squares is 
selected for decay of 2005 events. The rest observations denote by black triangles.
}
\label{saturation}
\end{figure}
\begin{figure}[ptbptbptb]
\includegraphics[scale=0.9,angle=0]{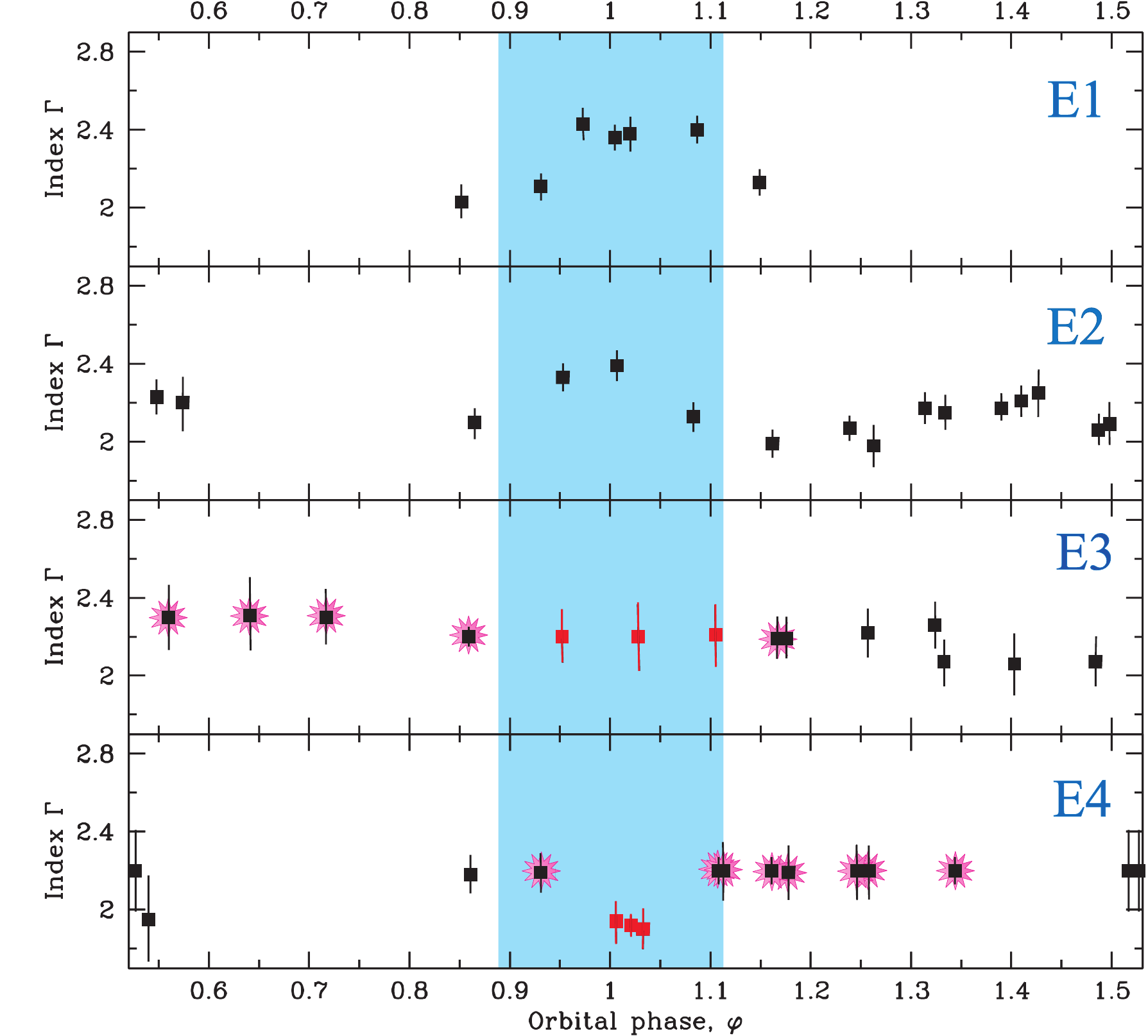}
\caption{
From top to bottom: Photon index $\Gamma$ versus orbital phase $\varphi$  
during MJD=50191-50194 (E1), 50897-50907 (E2), 52225-52238 (E3), 
53579-53588 (E4). Blue vertical strip marks an  interval of  the primary (optical) eclipse
\citep{gor98}. Points 
marked with rose oreol correspond to spectra fitted by the model which includes 
"high-temperature  BB" 
component (see also  Fig.~\ref{geometry} and Tables 4-6). 
Red points 
({\it for  two lower panels}) correspond to observations during  the primary eclipse 
when the  "high-temperature BB"  component is not detected, although it presents in spectral residual  before 
and after this eclipse.
}
\label{hard_BB_eclipse}
\end{figure}
\begin{figure}[ptbptbptb]
\includegraphics[scale=0.9,angle=0]{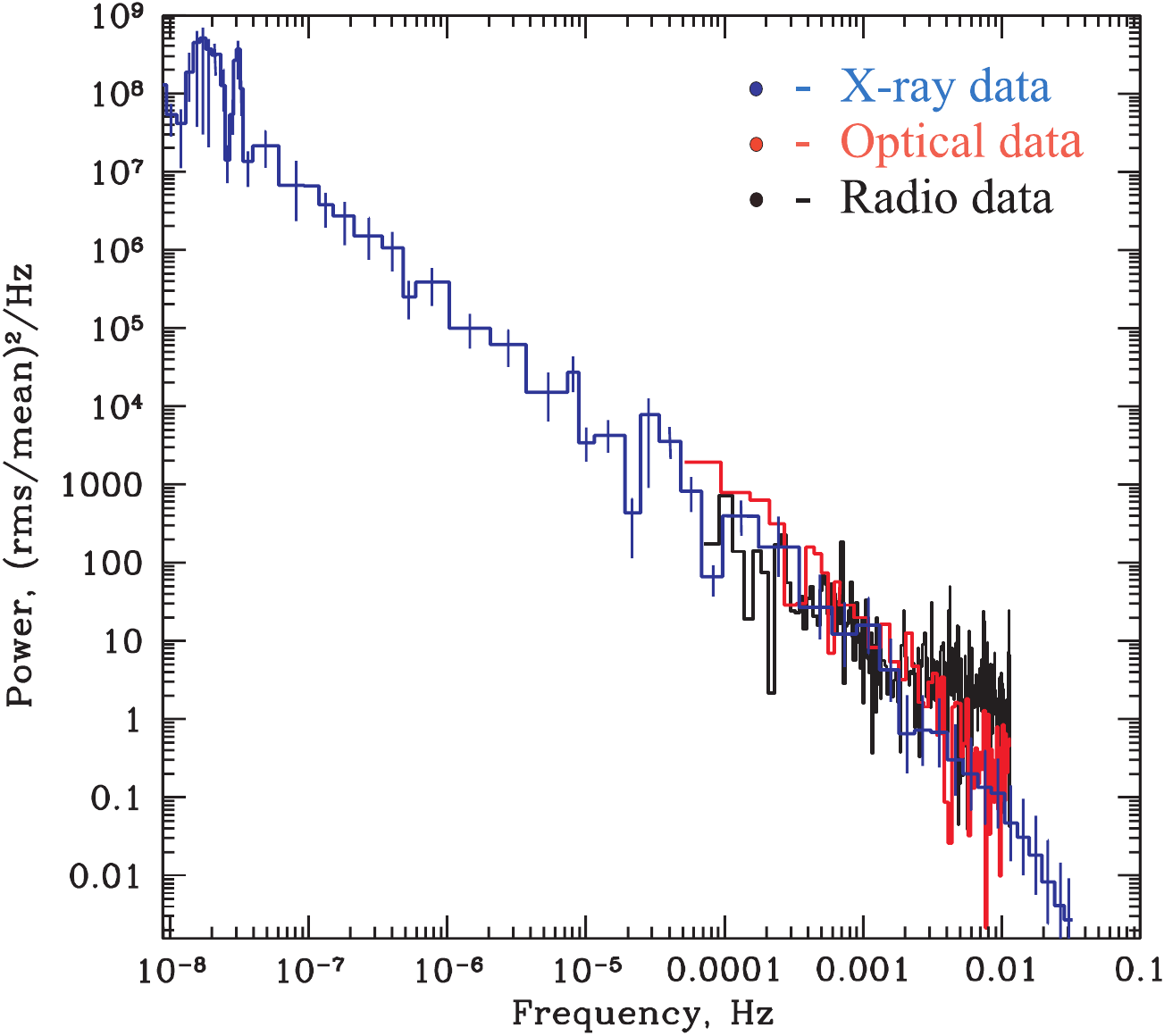}
\caption{
 Power spectrum of SS~433 in three energy bands: X-ray ({\it blue}), 2.25 GHz radio ({\it black}), 
V-optical ({\it red}).}
\label{pow_3band}
\end{figure}
\begin{figure}[ptbptbptb]
\includegraphics[scale=0.9,angle=0]{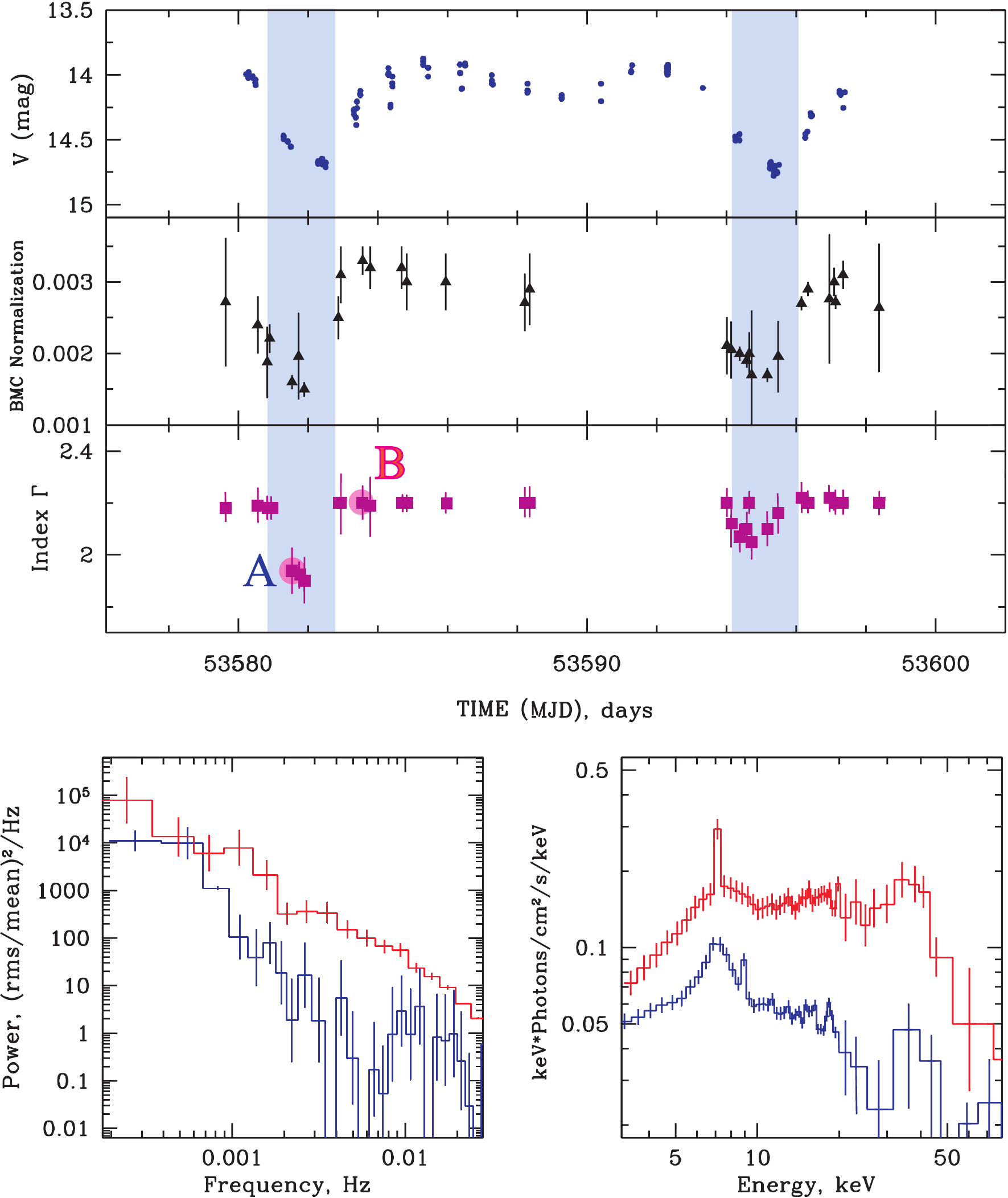}
\caption{{\it Top}:
From up to bottom:
Optical V-band light curve, 
BMC normalization and photon index $\Gamma$
for R6 set. Points A and B mark moments 53581 and 53585 (at eclipse 
and after eclipse) respectively.
Blue strips mark eclipsed intervals of light curve. 
{\it Bottom}: PDSs ({\it left panel}) are plotted along with energy spectral diagram 
$E\times F(E)$ ({\it right panel}) for observational points A ({\it91103-01-03-00, blue}) 
and B ({\it 91103-01-07-00, red}).
}
\label{spec_evol_R6}
\end{figure}
\begin{figure}[ptbptbptb]
\includegraphics[scale=1.,angle=0]{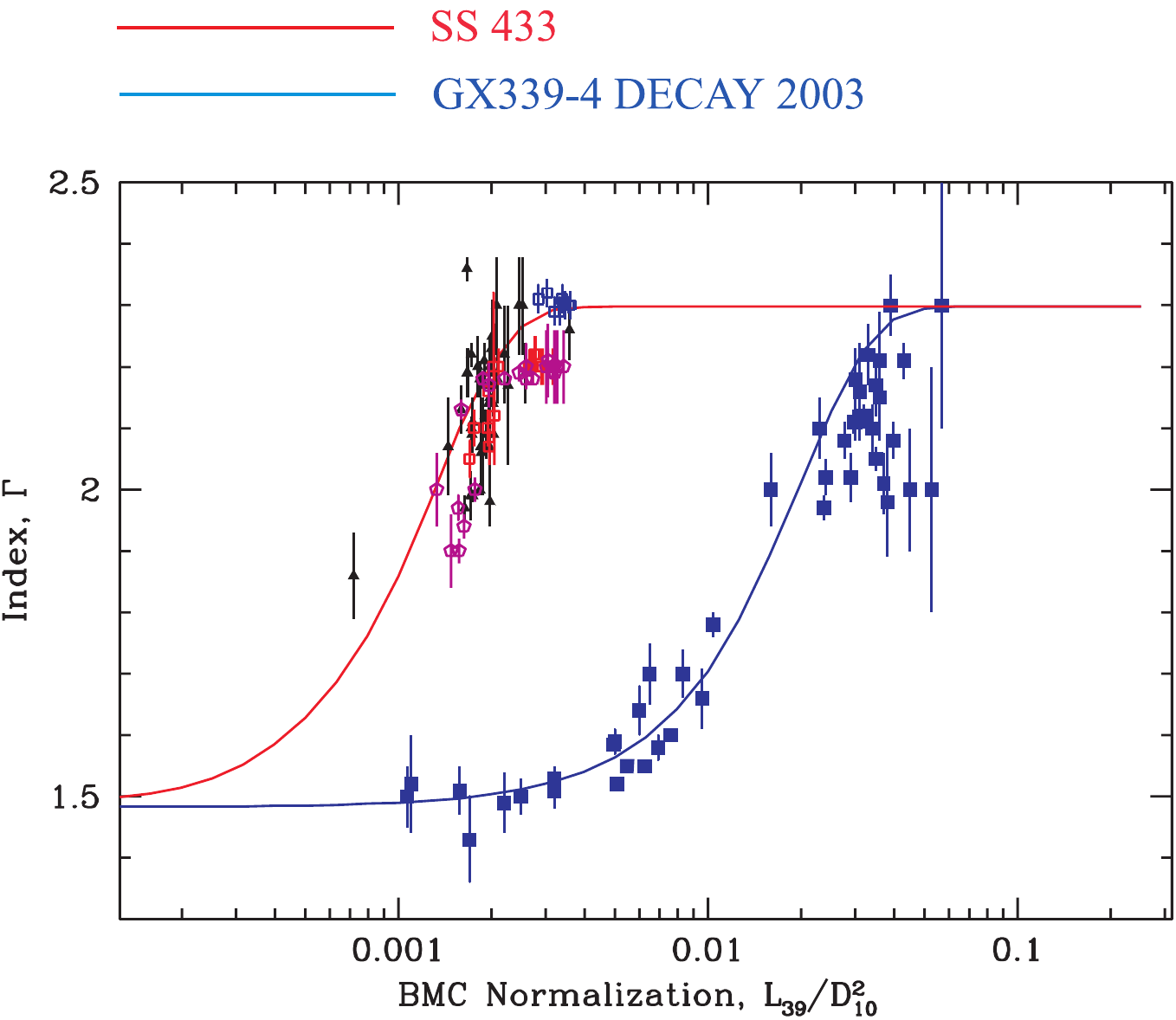}
\caption{
Scaling photon index versus normalization for SS~433 (with red line - target source) and 
GX 339-4 (with blue line - reference source)
}
\label{gx339_scal}
\end{figure}
\begin{figure}[ptbptbptb]
\includegraphics[scale=1.,angle=0]{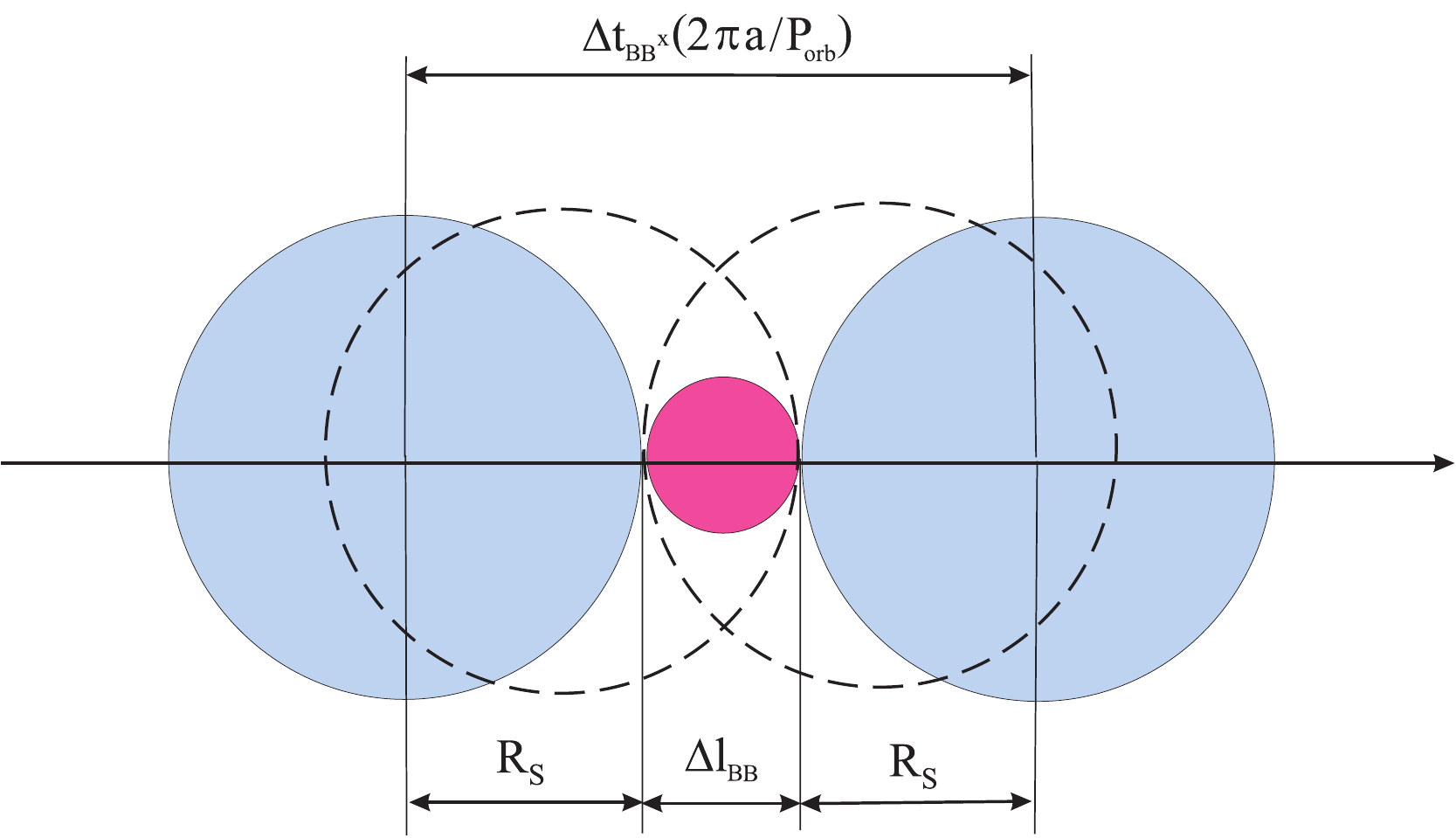}
\caption{
Eclipse of the inner X-ray emission region by  optical star.  
}
\label{eclipse_picture}
\end{figure}
\begin{figure}[ptbptbptb]
\includegraphics[scale=0.8,angle=0]{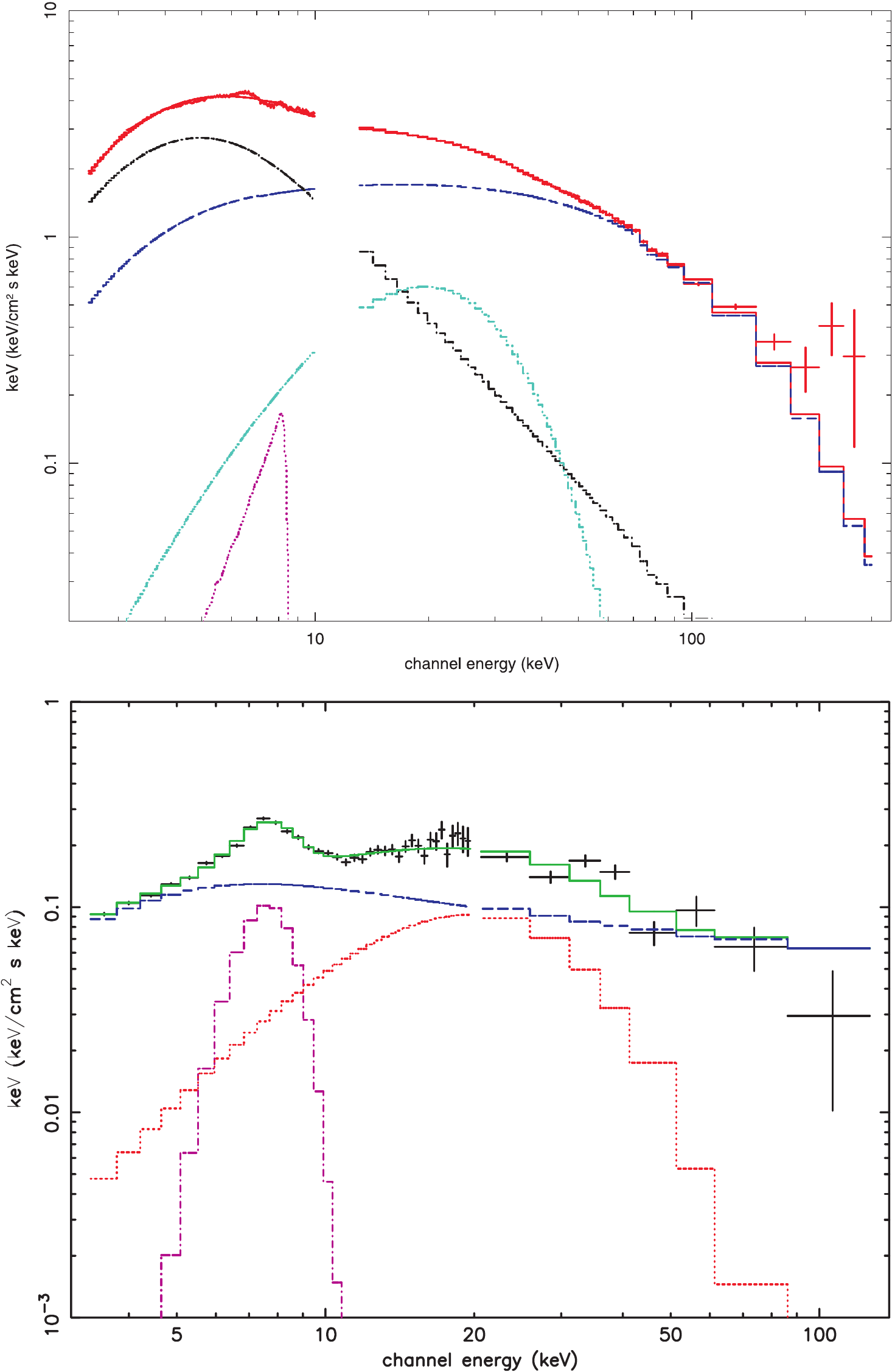}
\caption{ 
Broad-band BeppoSAX spectrum of GRS~1915+105 ({\it top}) and 
RXTE/PCA and IBIS/ISGRI/INTEGRAL spectrum of SS~433 ({\it bottom}) 
in $EF(E)$ units obtained during intermediate transition state of GRS~1915+105 
on April 21, 2000 (ID=209850011) and simultaneous {\it RXTE}/INTEGRAL 
observations of SS~433 for  outburst transition state 
on 
March 24 -- 27, 2004.
}
\label{sp_sax_int}
\end{figure}

{\it As   conclusions we formulate the following}.
 We analyze the state transition data from SS~433 collected using   \textit{RXTE}  observations. 
We examine the correlation between the photon index of the Comptonized spectral component 
and  its normalization which is presumably proportional to disk mass accretion rate  (see Fig.~\ref{saturation}). 
 We find  that broad-band energy spectra during all spectral states are well fit by  the {\tt XSPEC} {\it BMC}  model  for the continuum and by two (broad and narrow)
Gaussian line components. In addition to these  model components 
we also find a strong feature of "blackbody-like" bump which color 
temperature is in the range of 4-5 keV in 24 intermediate state (IS) spectra of 
SS~433.  

 Furthermore the application of our spectral model  to the SS~433 data allows  us to establish  the saturation of the photon index vs  
BMC normalization, which scales with disk mass accretion rate  at value around 2.3 (see Fig.~\ref{saturation}). 
In addition, an application of the scaling method (see ST09) allows us to estimate a lower limit of compact object  mass in SS~433 ($M_x >1.8 M_{\odot}$).

A high value  of the Comptonized 
emission fraction $f$ obtained, in the framework of BMC model,   gives us  a strong  evidence of significant reprocessing  of X-ray emission of the disk photons in SS 433 
which  is also  in agreement with power density spectra which reveal a pure power law (so called ``red noise''). 

It is important to emphasize that the index saturation effect in SS~433 now seen,  is a BH signature and was recently found in  a number of BHC sources (ST09).  Moreover  the detection of the so called  ``high temperature  BB-like''  bump'' (which could be  a gravitationally redshifted annihilation line) in SS~433  is  also   found  
 in GRS 1915+105 data  by   different  space missions  
 ({\it RXTE}, {\it BeppoSAX} and INTEGRAL). 


We acknowledge  Vitalij Goranskij and Tatyana Irsmambetova who kindly provide us optical data. 
We also thank  {\it RXTE},  INTEGRAL team and also Sergej Trushkin for providing  us  
X-ray and  radio data correspondingly. We appreciate comments and thorough editing of the paper 
made by Chris Shrader.  We are very grateful the referee for his/her valuable comments and corrections of the content of the paper.


\newpage

\begin{deluxetable}{llllll}
\tablewidth{0in}
\tabletypesize{\scriptsize}
    \tablecaption{The list of sets (groups) of RXTE observation of SS~433} 
    \renewcommand{\arraystretch}{1.2}
\tablehead{Number of set  & Dates, MJD & RXTE Proposal ID&  UT Dates  & Type of Light Curve & Ref.}
 \startdata
R1  &    50191-50194 & 10127        & Apr. 18 -- 21, 1996 & & 1 \\
R2  &    50868-50907 & 20102, 30273 & Feb. 24 -- Apr. 4, 1998 & outburst & 1,2 \\
R3  &    52222-52238 & 60058 & Nov. 9 -- 11, 2001 & outburst & 1  \\
R4  &    52544.46-52544.74, 52913-52914 & 70416, 80429 & Apr. 18 -- 21, 2002& & 1 \\
R5  &    53076-53092 & 90401        & Mar. 12 -- 28, 2004 & outburst & 1, 2  \\
R6  &    53239-53610 &  90401, 91103, 91092 & Jul.28 -- Aug. 28, 2005 & outburst decay & this work \\
R7  &    54085-54096 & 92424 & Dec. 17 -- 27, 2006 & outburst decay & this work \\
      \hline
      \enddata
    \label{tab:par_bbody}
References: 
(1) \cite{Fil06},  (2) \cite{nandi05}.
\end{deluxetable}

\bigskip


\begin{deluxetable}{cccc}
\tablewidth{0in}
\tabletypesize{\scriptsize}
    \tablecaption{The list of GRS~1915+105 observations  used in analysis.}
    \renewcommand{\arraystretch}{1.2}
\tablehead{
Satellite&Obs. ID& Start time (UT)  & End time (UT)}
\startdata
BeppoSAX&209850011& 2000 Apr. 21 08:55:30 & 2000 Apr. 21 15:16:47 \\
      \enddata
   \label{tab:table}
\end{deluxetable}

\newpage
\bigskip
\begin{deluxetable}{ccccccccccccccl}
\rotate
\tablewidth{0in}
\tabletypesize{\scriptsize}
    \tablecaption{Best-fit parameters of spectral analysis of PCA and HEXTE
observation of SS~433 in 3-150~keV energy range$^{\dagger}$.
Parameter errors correspond to 1$\sigma$ confidence level.}
    \renewcommand{\arraystretch}{1.2}
\tablehead{Observational & MJD,  & $\phi$ & $\psi$ & $\alpha=$  & log(A)$^{\dagger\dagger}$ & N$_{bmc}^{\dagger\dagger\dagger}$, & 
E$_{line1}$, &$\sigma_{line1}$&   $N_{line1}$ & E$_{line2}$, &N$_{line2}^{\dagger\dagger\dagger}$& $\chi^2_{red}$ (d.o.f.)& F$_1$/F$_2^{\dagger\dagger\dagger\dagger}$ &Rem\\
ID               & day  & & &   $\Gamma-1$          &           &$L_{39}/d^2_{10}$& keV & keV   &  &  keV        &  & & &$^{(a)}$ }
 \startdata
10127-01-01-00  &   50191.10 &   0.852 &  0.1820 & 1.03(2) & 2.00   &3.20(3) & 6.90(9) &  0.66(1) & 3.74(6)&  8.70(4) &0.28(4) & 1.17 (78) &  6.06/4.61 &  \\
10127-01-02-00  &   50192.10 &   0.931 &  0.1882 & 1.11(1) & 2.00   &2.23(3) & 6.87(1) &  0.73(2) & 3.1(1)&  8.83(5) &0.21(3) & 1.27 (78) &  4.40/3.13 & ecl \\
10127-01-03-00  &   50192.71 &   0.973 &  0.1911 & 1.43(3) & 2.00   &2.01(9) & 6.89(1) &  0.64(2) & 2.2(1)&  8.64(7) &0.19(4) & 0.77 (78) &  3.07/1.46 & ecl \\
10127-01-04-00  &   50193.06 &   0.005 &  0.1941 & 1.36(2) & 2.00   &1.87(2) & 6.88(1) &  0.69(2) & 2.3(1)&  8.75(6) &0.19(4) & 1.14 (78) &  2.99/1.51 & ecl \\
10127-01-05-00  &   50193.27 &   0.020 &  0.1950 & 1.38(3) & 2.00   &1.95(5) & 6.86(2) &  0.67(5) & 2.2(2)&  8.6(1)  &0.22(9) & 0.99 (78) &  3.25/1.87 & ecl \\
10127-01-06-00  &   50193.77 &   0.087 &  0.1980 & 1.40(2) & 2.00   &2.03(2) & 6.89(1) &  0.63(2) & 2.3(1)&  8.65(6)  &0.21(4) & 1.21 (78) &  3.18/1.58 & ecl \\
10127-01-07-00  &   50194.91 &   0.149 &  0.2051 & 1.13(1) & 2.00   &2.83(9) & 6.88(7) &  0.61(1) & 3.06(8)&  8.63(4)  &0.2(2) & 1.16 (78) &  5.15/3.64 &  \\
20102-01-01-00  &   50868.77 &   0.642 &  0.3609 & 0.97(2) & 2.00   &1.64(2) & 6.59(1) &  0.34(2) & 1.68(7)&  7.95(6) &0.23(4) & 1.01 (77) &  3.30/2.73 & \\
20102-01-02-00  &   50870.24 &   0.754 &  0.3700 & 1.00(1) & 2.00   &1.84(2) & 6.54(1) &  0.29(2) & 1.80(8)&  7.71(3) &0.39(5) & 1.57 (77) &  3.42/2.97 &  \\
20102-01-03-00  &   50871.91 &   0.895 &  0.3810 & 1.15(2) & 2.00   &2.14(5) & 6.56(1) &  0.26(3) & 1.9(1)&  7.70(4) &0.47(7) & 1.28 (78) &  3.79/2.64 & ecl \\
20102-01-04-00  &   50873.18 &   0.991 &  0.3850 & 1.36(3) & 2.00   &1.67(7) & 6.58(1) &  0.36(2) & 1.82(8)&  7.95(3) &0.36(4) & 1.02 (78) &  3.79/2.64 & ecl \\
20102-01-05-00  &   50874.92 &   0.111 &  0.3988 & 1.19(4) & 2.00   &1.68(6) & 6.57(2) &  0.33(5) & 2.1(2) &  7.9(1)  &0.3(1)  & 0.85 (77) &  2.74/1.84 &  \\ 
20102-01-06-00  &   50876.72 &   0.250 &  0.4099 & 1.19(4) & 2.00   &1.67(6) & 6.57(2) &  0.33(5) & 2.1(2) &  7.9(1)  &0.3(1)  & 0.85 (77) &  2.73/1.80 &  \\
20102-02-01-02  &   50876.79 &   0.255 &  0.4104 & 1.20(5) & 2.00   &1.81(6) & 6.58(3) &  0.36(6) & 2.2(2) &  7.8(3)  &0.03(1) & 0.93 (77) &  2.90/1.94 &  \\
20102-02-01-03  &   50877.58 &   0.315 &  0.4152 & 1.11(4) & 2.00   &1.74(6) & 6.59(2) &  0.43(4) & 2.4(2) &  8.01(9) &0.4(1)  & 0.98 (77) &  3.01/2.21 &  \\
20102-02-01-04  &   50877.64 &   0.320 &  0.4156 & 1.36(2) & 2.00   &1.67(2) & 6.58(1) &  0.36(2) & 1.82(8)&  7.95(3) &0.36(4) & 1.01 (77) &  2.60/1.32 &  \\
20102-02-01-05  &   50877.71 &   0.325 &  0.4160 & 1.18(3) & 2.00   &1.84(4) & 6.61(2) &  0.39(4) & 2.23(7)&  7.96(8) &0.4(1)  & 0.70 (77) &  3.01/2.10 &  \\
20102-02-01-01  &   50877.79 &   0.331 &  0.4165 & 1.14(1) & 2.00   &1.96(2) & 6.63(1) &  0.42(1) & 2.37(8)&  8.01(4) &0.30(5) & 0.94 (77) &  3.34/2.45 &  \\
20102-02-01-06  &   50878.00 &   0.347 &  0.4178 & 1.09(1) & 2.00   &2.03(1) & 6.637(6)&  0.42(1) & 2.53(6)&  8.06(3) &0.34(3) & 1.42 (77) &  3.51/2.75 &  \\
20102-02-01-07  &   50878.58 &   0.391 &  0.4214 & 1.18(4) & 2.00   &2.09(6) & 6.68(2) &  0.46(4) & 2.5(2) &  8.1(3)  &0.2(1)  & 0.84 (77) &  3.41/2.42 &  \\
20102-02-01-000 &   50878.64 &   0.396 &  0.4218 & 1.14(2) & 2.00   &2.0(1)  & 6.64(8) &  0.44(1) & 2.41(7)&  8.6(1)  &0.04(2) & 0.63 (77) &  2.84/1.89 &  \\
20102-02-01-00  &   50878.97 &   0.421 &  0.4238 & 1.12(1) & 2.00   &1.9(1)  & 6.67(8) &  0.46(1) & 2.44(7)&  8.20(4) &0.2(3)  & 0.84 (77) &  3.52/2.42 &  \\
20102-02-01-11  &   50879.16 &   0.436 &  0.4249 & 1.09(3) & 2.00   &1.73(5) & 6.62(1) &  0.45(3) & 2.24(4)&  8.02(6) &0.30(7) & 1.34 (77) &  3.02/2.32 &  \\
20102-02-01-08  &   50879.37 &   0.452 &  0.4263 & 0.99(4) & 2.00   &1.71(9) & 6.60(2) &  0.50(5) & 2.6(2) &  8.1(1)  &0.3(1)  & 0.68 (77) &  3.33/2.81 &  \\
20102-02-01-09  &   50879.51 &   0.463 &  0.4271 & 1.13(4) & 2.00   &1.60(6) & 6.66(1) &  0.60(3) & 2.8(2) &  8.9(1)  &0.19(6) & 0.89 (77) &  3.04/1.92 &  \\
20102-02-01-10  &   50879.57 &   0.467 &  0.4275 & 1.22(2) & 2.00   &1.72(3) & 6.66(1) &  0.46(2) & 2.4(1) &  8.22(7) &0.26(5) & 0.79 (77) &  2.85/1.82 &  \\
20102-01-07-00  &   50889.73 &   0.244 &  0.4902 & 0.86(7) & 0.36(9)&0.72(3) & 6.70(3) &  0.59(6) & 0.44(7)&  8.6(1)  &0.04(2) & 1.03 (76) &  1.22/1.25 &  \\
20102-02-02-01  &   50897.85 &   0.865 &  0.5402 & 1.10(2) & 2.00   &1.72(3) & 6.16(1) &  0.45(3) & 1.9(1) &  8.05(5) &0.32(5) & 0.79 (77) &  3.04/2.24 &  \\
20102-02-02-00  &   50898.78 &   0.953 &  0.546  & 1.33(1) & 2.00   &1.85(2) & 6.60(1) &  0.35(2) & 1.90(8) & 7.92(3) &0.38(4) & 1.44 (77) &  3.00/1.40 & ecl \\
30273-01-01-00  &   50899.72 &   0.007 &  0.551 & 1.39(1) & 2.00   &1.58(1) & 6.58(1) &  0.35(1) & 1.81(4)&  7.92(3) &0.39(4) & 1.38 (77) &  2.49/1.21 &  ecl\\
30273-01-01-01  &   50900.73 &   0.083 &  0.558 & 1.13(1) & 2.00   &1.72(3) & 6.56(1) &  0.34(2) & 1.82(8)&  7.83(3) &0.29(4) & 1.40 (77) &  3.16/2.17 &  ecl\\
30273-01-02-00  &   50901.74 &   0.162 &  0.5642 & 0.99(1) & 2.00   &1.74(3) & 6.60(1) &  0.41(2) & 2.02(8)&  7.92(3) &0.29(4) & 1.32 (77) &  3.32/2.85 &  \\
30273-01-02-010 &   50902.74 &   0.239 &  0.5704 & 1.07(1) & 2.00   &1.93(2) & 6.643(8)&  0.47(1) & 2.32(7)&  8.12(3) &0.31(4) & 1.20 (77) &  3.52/2.73 &  \\
30273-01-02-01  &   50903.06 &   0.263 &  0.5724 & 0.98(4) & 2.00   &1.98(9) & 6.67(2) &  0.55(5) & 2.6(2) &  8.3(1)  &0.3(1)  & 0.80 (77) &  3.81/3.21 &  \\
30273-01-03-000 &   50903.73 &   0.314 &  0.5765 & 1.17(1) & 2.00   &2.01(2) & 6.616(8)&  0.40(1) & 2.32(7)&  7.98(3) &0.36(4) & 1.02 (77) &  3.35/2.34 &  \\
30273-01-03-00  &   50903.99 &   0.334 &  0.5781 & 1.15(3) & 2.00   &1.98(5) & 6.63(1) &  0.42(3) & 2.4(1) &  8.13(8) &0.33(8) & 0.98 (77) &  3.32/2.31 &  \\
30273-01-03-010 &   50904.72 &   0.390 &  0.5826 & 1.17(1) & 2.00   &1.90(2) & 6.565(8)&  0.33(7) & 2.09(7)&  7.80(2) &0.47(4) & 0.98 (77) &  3.31/2.30 &  \\
30273-01-03-01  &   50904.99 &   0.410 &  0.5843 & 1.21(3) & 2.00   &1.90(4) & 6.59(2) &  0.34(4) & 2.1(1) &  7.88(7) &0.45(1) & 0.76 (77) &  3.12/2.04 &  \\      
30273-01-03-02  &   50905.20 &   0.427 &  0.5856 & 1.25(6) & 2.00   &2.01(6) & 6.58(3) &  0.35(7) & 2.01(6)&  7.8(1)  &0.4(1)  & 1.29 (77) &  3.21/2.01 &  \\
30273-01-05-01  &   50906.00 &   0.487 &  0.5905 & 1.06(2) & 2.00   &1.86(4) & 6.55(1) &  0.35(3) & 2.2(1) &  7.94(6) &0.36(7) & 1.01 (77) &  3.32/2.61 &  \\
30273-01-05-03  &   50906.13 &   0.498 &  0.5913 & 1.09(4) & 2.00   &1.93(6) & 6.54(2) &  0.38(5) & 2.2(2) &  7.8(1)  &0.38(5) & 1.09 (77) &  3.42/2.64 &  \\
30273-01-05-00  &   50906.79 &   0.548 &  0.5953 & 1.23(1) & 2.00   &2.00(2) & 6.533(8)&  0.28(1) & 2.04(7)&  7.7(2)  &0.47(4) & 0.95 (77) &  3.24/2.12 &  \\
30273-01-05-02  &   50907.13 &   0.574 &  0.5975 & 1.20(5) & 2.00   &1.98(6) & 6.54(2) &  0.34(5) & 2.2(2) &  7.9(1)  &0.4(1)  & 0.81 (77) &  3.21/2.21 &  \\
60058-01-01-00  &   52222.29 &   0.105 &  0.7082 & 1.07(7) & 2.00   &1.97(7) & 6.67(4) &  0.42(7) & 1.8(2) &  7.9(1)  &0.27(7) & 0.79 (77) &  3.47/2.84 &  \\
60058-01-02-00  &   52223.22 &   0.176 &  0.7139 & 1.19(8) & 2.00   &2.04(6) & 6.95(4) &  0.43(7) & 1.3(2) &  7.9(1)  &0.3(1)  & 1.15 (77) &  3.22/2.53 &  \\
60058-01-03-00  &   52224.28 &   0.257 &  0.7205 & 1.22(8) & 2.00   &2.21(7) & 6.94(1) &  0.6(1)  & 1.3(3) &  6.6(1)  &0.5(3)  & 1.37 (77) &  3.42/2.61 &  \\
60058-01-04-00* &   52225.27 &   0.333 &  0.7266 & 1.07(8) &-0.5(4) &1.5(3)  & 7.01(1) &  0.72(4) & 2.7(3) &  9.4(1)  &0.17(9) & 1.09 (69) & 3.33/1.73  &  \\
60058-01-05-00* &   52226.19 &   0.403 &  0.7323 & 1.06(8) &-0.5(4) &1.9(3)  & 7.03(2) &  0.63(3) & 2.5(1) &  9.4(1)  &0.02(9) & 1.30 (69) & 3.91/1.95  &  \\
60058-01-06-00* &   52227.25 &   0.484 &  0.7388 & 1.07(5) &-0.1(1) &1.8(3)  & 7.05(1) &  0.74(5) & 2.6(3) &  9.4(1)  &0.10(9) & 1.01 (69) & 3.74/2.13  &  \\
60058-01-07-00* &   52228.24 &   0.560 &  0.7449 & 1.30(6) & 2.00   &2.46(4) & 6.90(1) &  0.52(3) & 2.1(1) &  8.9(1)  &0.15(8) & 1.08 (69) & 4.15/2.69  &  \\
60058-01-08-00* &   52229.30 &   0.641 &  0.7515 & 1.31(8) & 0.7(3) &2.5(1)  & 6.95(3) &  0.55(5) & 2.2(2) &  8.80(4) &0.25(8) & 1.20 (69) & 4.12/2.78  & \\
60058-01-09-00* &   52230.30 &   0.717 &  0.7576 & 1.30(5) & 0.5(2) &2.1(1)  & 6.92(4) &  0.73(6) & 2.18(3)&  8.69(4) &0.25(8) & 0.79 (69) & 3.46/2.17  &  \\
60058-01-10-00* &   52232.15 &   0.859 &  0.7690 & 1.2(1)  & 2.00   &2.27(3) & 6.96(3) &  0.86(4) & 3.2(3) &     -    &   -    & 1.2 (69) &  4.15/2.74  & \\
60058-01-11-00* &   52233.27 &   0.952 &  0.7758 & 1.20(8)  & 2.00   &2.06(1) & 6.97(4) &  0.63(7) & 1.9(2) & 8.7(2) & 0.2(1) & 1.2 (69) &  3.57/2.37  & ecl\\
60058-01-12-00 &   52234.33 &   0.028 &  0.7819 & 1.20(9)  & 2.00   &1.8(1) & 7.03(5) &  0.68(9) & 1.7(3) & 8.7(2) & 0.1(1) & 1.04 (77) &  3.16/2.07  & ecl\\
60058-01-13-00 &   52235.33 &   0.105 &  0.7881 & 1.21(8)  & 2.00   &2.60(8) & 6.95(3) &  0.55(5) & 2.6(8) & 9.0(4) & 0.2(1) & 1.17 (77) &  4.27/3.08  & ecl\\
60058-01-15-00* &   52236.18 &   0.167 &  0.7939 & 1.19(5) & 2.00   &2.58(2) & 6.99(2) &  0.67(4) & 2.6(2) &     -    &   -    & 1.09 (69) & 4.85/3.42  &  \\
60058-01-17-00  &   52238.23 &   0.324 &  0.8065 & 1.26(5) & 2.00   &3.57(7) & 7.06(3) &  0.69(4) & 2.5(2) &     -    &   -    & 1.59 (71) &  5.71/3.94 &  \\
70416-01-01-00  &   52544.46 &   0.732 &  0.6951 & 1.13(5) & 2.00   &1.60(7) & 6.64(4) &  0.2(1)  & 1.6(2) &  7.9(1)  &0.3(1)& 0.93 (72)& 2.99/2.04  & \\
70416-01-01-01  &   52544.66 &   0.747 &  0.6963 & 1.00(6) & 0.4(1) &1.7(1)  & 6.57(6) &  0.2(1)  & 1.2(2) &  7.7(1)  &0.4(1)& 0.81 (71)& 2.76/1.98  & \\
70416-01-01-02  &   52544.74 &   0.753 &  0.6968 & 0.97(6) & 0.7(2) &1.5(1)  & 6.61(6) &  0.3(1)  & 1.61(6)&  7.8(1)  &0.4(2)& 0.87 (72)& 2.92/2.23  & \\
80429-01-01-00  &   52913.69 &   0.963 &  0.9711 & 1.21(8) & 2.0    &1.8(1)  & 7.30(9) &  1.2(1)  & 2.3(5)&  -        & -     & 1.13 (72)& 3.36/2.03  & ecl\\
80429-01-01-01  &   52914.22 &   0.007 &  0.9761 & 1.22(5) & 2.0    &1.84(8)  & 7.42(5) &  1.8(2)  & 1.06(7)&  -        & -     & 1.03 (72)& 3.22/1.983  & ecl\\
90401-01-01-01* &   53076.78 &   0.423 &  0.9780 & 1.30(2) & 0.4(2) &3.4(1)  & 7.23(3) &  0.96(8) & 4.7(6) &  9.2(3)  &0.1(1)& 1.01 (69) & 8.59/6.54 &  \\
90401-01-01-03* &   53076.85 &   0.428 &  0.9784 & 1.30(2) & 0.3(1) &3.4(1)  & 7.15(2) &  0.80(4) & 4.4(8) &  9.3(2)  &0.26(8)& 1.03 (69) & 7.91/4.85 &  \\
90401-01-01-00* &   53077.77 &   0.498 &  0.9841 & 1.32(2) & 0.3(1) &3.03(6) & 7.06(1) &  0.75(2) & 4.3(2) &  9.08(5) &0.30(5)& 1.24 (69) & 7.72/4.75 &  \\
90401-01-01-02* &   53078.75 &   0.573 &  0.9901 & 1.30(1) & 2.00   &3.5(1)  & 7.02(3) &  0.87(4) & 4.1(4) &  8.9(2)  &0.1(1)& 1.35 (69) & 7.01/4.63 &  \\
90401-01-02-01* &   53089.07 &   0.362 &  0.0538 & 1.31(2) & 0.9(6) &3.4(1)  & 7.07(2) &  0.97(4) & 5.3(3) &     -    &   -  &1.17 (69) &     5.25/3.01 &  \\
90401-01-02-00* &   53089.28 &   0.378 &  0.0550 & 1.30(2) & 1.8(1) &3.6(1)  & 7.12(2) &  0.80(3) & 4.3(1) &  9.2(1)  &0.80(3)& 1.34 (69) & 8.01/5.47 &   \\
90401-01-03-01* &   53091.04 &   0.512 &  0.0659 & 1.29(2) & 0.16(2)& 3.2(1) & 7.02(3) &  0.70(7) & 3.7(4) &  8.8(1)  &0.4(1) &1.14 (69) & 5.64/3.58 &   \\
90401-01-03-02* &   53091.82 &   0.572 &  0.0707 & 1.29(2) & 0.4(1) & 3.3(1) & 7.04(2) &  0.73(4) & 3.8(3) &  9.1(4)  &0.21(9) &1.08 (69) & 5.95/3.51 &   \\
90401-01-03-00* &   53092.09 &   0.593 &  0.0724 & 1.31(2) & 2.00   &2.8(1)  & 7.10(2) &  0.90(4) & 4.2(3) &     -    &   -  &1.19 (69) &     5.93/3.62 & \\
90401-01-04-01  &   53239.49 &   0.860 &  0.9814 & 1.2(1)  & 2.00   & 3.42(8)& 7.33(3) &  0.71(4) & 2.4(1) &     -    &   -  &1.50 (73)& 5.62/3.98      & \\
90401-01-05-01  &   53361.71 &   0.202 &  0.7352 & 1.00(6) & 2.00   & 1.24(9)& 6.58(2) &  0.42(4) & 1.8(1) &     -    &   -  &1.15 (73)& 2.89/2.12      & \\
90401-01-06-00  &   53363.75 &   0.359 &  0.7478 & 1.09(2) & 2.00   & 1.46(9)& 6.60(3) &  0.31(4) & 1.8(1) &  8.3(1)  &0.27(9)& 1.19 (73)& 2.87/1.91  & \\
90401-01-06-01  &   53366.13 &   0.540 &  0.7624 & 0.95(7) & 2.00   & 1.71(6)& 6.62(3) &  0.33(7) & 2.4(2) &  8.3(1)  &0.4(1)& 1.15 (71)& 3.45/2.96  & \\
91103-01-01-00  &   53579.63 &   0.861 &  0.0791 & 1.18(3) & 2.00   & 2.72(9)& 7.09(2) &  0.91(2) & 3.9(3) &     -    &   -  &1.17 (71)& 2.78/1.63      & \\
91103-01-09-00  &   53580.55 &  0.931 &  0.0848 & 1.19(4) &  2.00  &  2.4(4)  &  7.22(4)  &   0.87(6)  &  2.1(2)  &   -      &   -  &  0.97(74) &  4.20/2.57 & ecl\\
91103-01-02-01  &   53580.83 &  0.952 &  0.0865 & 1.18(3) &  2.00  &  1.88(5) &  7.1(1)   &   0.78(2)  &  1.7(5)  &   8.7(1) &   0.2(2) & 1.18(73) &  3.38/2.13 &ecl \\
91103-01-02-00* &   53580.90 &  0.958 &  0.0870 & 1.18(3) &  0.2(2) &   2.21(2)&  6.70(2) &   0.27(3)  &  0.6(3)  &   8.0(2) &   0.4(1) & 0.97(69) &  2.97/1.74 &ecl \\
91103-01-03-00  &   53581.53 &  0.006 &  0.0908 & 0.94(6) &  0.58(5) &  1.6(1) &  6.96(3) &   0.99(3)  &  1.61(6) &   9.3(1) &   0.10(8)& 1.02(73) &  2.79/2.30 &ecl, "A" \\
91103-01-04-01  &   53581.73 &  0.021 &  0.0920 & 0.92(2) &  0.6(1) &   1.95(1)&  7.5(1)  &   0.9(1)   &  1.1(4)  &   6.9(1) &   0.4(2) & 1.04(71) &  2.91/1.90 &ecl \\
91103-01-04-00  &   53581.89 &  0.033 &  0.0930 & 0.90(6) &  0.6(1) &   1.5(1) &  6.8(1)  &   1.29(1)  &  3.0(5)  &    -     &   -  &  0.85(73) &  3.22/2.52 &ecl \\
91103-01-05-00* &   53582.87 &  0.108 &  0.0991 & 1.2(1)  & 0.01(4)& 2.5(3) & 7.08(6) &  0.75(8) & 2.8(6) &     -    &   -  &0.97 (69)& 3.90/2.13 & \\
91103-01-05-01* &   53582.94 &  0.113 &  0.0995 & 1.20(6) & 0.9(1) & 3.1(4) & 7.02(6) &  0.75(9) & 3.0(6) &     -    &   -  &1.07 (69)& 5.41/3.34 & \\
91103-01-07-00* &   53583.56 &  0.161 &  0.1034 & 1.2(1)  & 2.00   & 3.3(2) & 7.03(3) &  0.75(6) & 2.8(3) &  7.1(4)  &0.9(2)&1.04 (69)& 6.30/3.88 & "B"\\
91103-01-06-00* &   53583.78 &  0.178 &  0.1047 & 1.19(2) & 2.00   & 3.2(3) & 7.08(5) &  0.90(9) & 3.0(4) &  7.0(1)  &0.05(3)&1.25 (69)& 5.56/3.44 & \\
91103-01-06-01* &   53584.68 &  0.246 &  0.1102 & 1.2(1)  & 2.00   & 3.2(3) & 7.06(3) &  0.75(5) & 3.5(3) &  7.01(8) &0.6(3) &1.03 (69)& 6.79/4.01 & \\
91103-01-08-00* &   53584.83 &  0.258 &  0.1112 & 1.2(1)  & 2.00   & 3.0(4) & 7.12(9) &  0.98(6) & 4.1(3) &  7.0(1)  &0.5(3) &1.03 (69)& 6.56/4.29 & \\
91103-01-09-01* &   53585.95 &  0.344 &  0.1181 & 1.2(1)  & 2.00   & 3.0(4) & 6.87(6) &  0.84(9) & 3.9(7) &     -    &   - & 1.05 (69)& 6.54/4.34     & \\
91092-01-01-00* &   53588.22 &  0.517 &  0.1321 & 1.20(3) & 0.6(3) & 2.71(4)& 6.89(3) &  0.75(6) & 3.7(4) &  9.1(1)  &0.2(1)&0.96 (69) & 4.36/2.56 & \\
91092-01-02-00* &   53588.35 &  0.527 &  0.1329 & 1.20(3) & 0.14(9)& 2.90(5)& 6.88(1) &  0.50(2) & 2.8(1) &     -    &   -  &1.17 (69) &    5.93/3.58  & \\
91092-02-01-02  &   53594.01 &  0.960 &  0.1678 & 1.20(3) &  0.57(4) &  2.11(4) &  6.90(5) &   0.46(7)  &  1.6(1)  &   8.3(1) &   0.3(1)  & 1.08(73) &  3.22/1.92  & ecl\\
91092-02-01-04  &   53594.14 &  0.970 &  0.1686 & 1.12(8) &  0.45(6) &  2.05(4) &  6.98(6) &   0.52(6)  &  1.5(2)  &   8.6(1) &   0.19(9) & 1.08(73) &  3.13/1.98  & ecl\\
91092-02-01-00  &   53594.38 &  0.988 &  0.1701 & 1.07(3) &  0.35(2) &  2.0(1)  &  6.89(2) &   0.38(2)  &  1.3(1)  &   8.37(8)&   0.34(8) &1.12(73)  &  2.98/1.95  & ecl\\
91092-02-01-01  &   53594.58 &  0.004 &  0.1713 & 1.10(3) &  0.5(1)  &  1.9(1)  &  6.84(3) &   0.36(7)  &  1.3(2)  &   7.9(7) &   0.3(2)  &0.70(73)  &  3.10/2.06  & ecl\\
91092-02-07-01  &   53594.65 &  0.008 &  0.1717 & 1.2(1)  &  0.31(4) &  2.0(3)  &  6.70(6) &   0.38(8)  &  1.1(1)  &   8.3(2) &   0.2(1)  &1.13(71)  &   2.78/1.63 & ecl\\
91092-02-01-03  &   53594.72 &  0.014 &  0.1722 & 1.05(3) &  0.46(7) &  1.70(9) &  6.88(5) &   0.60(5)  &  1.6(3)  &   8.6(1) &   0.3(1)  &1.04(72)  &  2.87/1.96  & ecl\\
91092-02-02-00  &   53595.17 &  0.048 &  0.1749 & 1.10(3) &  0.4(3)  &  1.7(1)  &  6.90(3) &   0.65(6)  &  1.71(2) &   8.72(9)&   0.20(7) & 1.11(72) &  3.04/2.02  & ecl\\
91092-02-03-00  &   53595.49 &  0.073 &  0.1769 & 1.16(1) &  2.00(0) &  1.96(5) &  6.81(1) &   0.57(1)  &  1.9(2)  &   8.5(1) &   0.03(1) &1.28(72)  &  3.62/2.32  & ecl\\
91092-02-04-00G*&   53596.15 &  0.123 &  0.1810 & 1.22(3) & 0.6(4) & 2.7(1) & 6.81(2) &  0.48(5) & 1.7(1) &  8.37(5) &0.40(7) &1.46 (69) & 5.63/3.54 &   \\
91092-02-05-00* &   53596.34 &  0.138 &  0.1822 & 1.20(3) & 0.5(3) & 2.9(1) & 6.83(1) &  0.46(4) & 2.6(2) &  8.44(8) &1.2(4)  &1.46 (69) & 5.29/3.57 &   \\
91092-02-06-01* &   53596.96 &  0.185 &  0.1860 & 1.22(3) & 0.6(2) & 2.76(9)& 6.90(2) &  0.53(3) & 2.5(1) &  8.50(8) &0.32(7) &1.04 (69) & 5.14/3.26 &   \\
91092-02-06-02* &   53597.09 &  0.195 &  0.1868 & 1.20(3) & 2.00   & 3.0(2) & 6.93(4) &  0.54(8) & 2.2(2) &  8.7(3)  &0.2(1)  &1.02 (69) & 5.25/3.58 & \\
91092-02-06-00* &   53597.13 &  0.198 &  0.1871 & 1.20(3) & 1.1(4) & 2.72(1)& 6.80(2) &  0.50(3) & 2.5(1) &  8.30(8) &0.43(8) &1.05 (69)& 5.52/3.53  & \\
91092-02-07-00  &   53597.34 &  0.214 &  0.1883 & 1.20(3) & 2.00   & 3.1(2) & 6.88(2) &  0.46(4) & 2.2(2) &  8.5(1)  &0.17(8) &1.10 (72) & 5.41/3.73 & \\
91092-02-08-00*  &   53598.38 &  0.293 &  0.1947 & 1.20(1) & 0.1(1) & 2.64(9)& 6.80(6) &  0.53(2) & 1.6(3) &  8.3(2)  &0.2(1) &1.13 (71) & 3.358/1.71 & \\
91103-01-10-00  &   53610.77 &  0.241 &  0.2712 & 1.18(4) & 2.00   & 2.59(3)& 6.63(2) &  0.40(4) & 2.3(1) &     -    &   - & 1.08 (73)& 4.12/2.75       & \\
92424-01-02-05  &   54085.94 &  0.563 &  0.2016 & 1.18(7) & 2.00   & 2.73(5)& 6.84(3) &  0.59(5) & 2.8(2) &     -    &   - & 1.17 (73)& 4.51/3.28       & \\
92424-01-01-00  &   54086.99 &  0.643 &  0.2081 & 1.19(4) & 2.00   & 3.11(7)& 6.70(3) &  0.51(8) & 2.5(3) &  8.2(1)  &0.2(2)& 1.10 (71)& 5.42/3.67   & \\
92424-01-01-01  &   54087.97 &  0.718 &  0.2141 & 1.16(7) & 2.00   & 2.95(7)& 6.68(2) &  0.31(3) & 2.2(3) &  7.9(2)  &0.3(2)& 1.33 (71)& 5.57/4.26   & \\
92424-01-01-02  &   54088.95 &  0.793 &  0.2202 & 1.21(5) & 2.00   & 3.13(5)& 6.66(1) &  0.29(7) & 2.0(5) &  7.8(2)  &0.6(4)& 1.43 (71)& 5.36/3.61   & \\
92424-01-01-03  &   54089.92 &  0.867 &  0.2261 & 1.21(7) & 2.00   & 2.64(3)& 6.67(2) &  0.44(4) & 2.4(1) &     -    &   - & 1.43 (73)& 4.59/2.99       & \\
92424-01-02-02  &   54093.85 &  0.167 &  0.2504 & 1.16(3) & 2.00   & 2.31(4)& 6.53(2) &  0.16(6) & 2.3(2) &  7.9(1)  &0.4(1)& 1.22 (71)& 3.94/2.79   & \\
92424-01-02-03  &   54094.91 &  0.249 &  0.2569 & 1.16(3) & 2.00   & 2.10(4)& 6.51(3) &  0.32(8) & 2.2(2) &  7.9(1)  &0.4(1)& 1.30 (71)& 3.85/2.48   & \\
92424-01-02-04  &   54095.91 &  0.325 &  0.2630 & 1.16(3) & 2.00   & 2.34(6)& 6.53(5) &  0.23(3) & 1.9(3) &  7.8(3)  &0.3(2)& 0.93 (71)& 4.01/2.67   & \\
92424-01-02-05  &   54096.73 &  0.388 &  0.2681 & 1.16(3) & 2.00   & 2.13(5)& 6.48(2) &  0.05(3) & 1.8(3) &  7.6(1)  &0.6(2)& 0.86 (71)& 3.86/2.54   & \\
      \enddata
    \label{tab:fit_table}
$^\dagger$ The spectral model is  $wabs*(bmc + Gaussian1 + Gaussian2)$,
$^{\dagger\dagger}$when parameter $log(A)>1$, then it is fixed at  2.0 (see comments in the text), 
$^{\dagger\dagger\dagger}$ normalization parameter of Gaussian1 and Gaussian2 components are in units of $10^{-3} (10^{-10}$ ergs/s/cm$^2)$,  
$\sigma_{line2}$ of Gaussian2 component is fixed at 0.01 keV (see comments in the text), 
$^{\dagger\dagger\dagger\dagger}$spectral fluxes (F$_1$/F$_2$) in the 3 -- 60/13 -- 150 energy ranges, correspondingly, in units of $10^{-10}$ ergs/s/cm$^2$,
* this observations are  fitted with $wabs*(bmc+Gaussian1+Gaussian2+bbody)$ model, see values of the best-fit BB color temperature and EW in Table 4, 5 and 6,
$^{(a)}$term ``ecl'' marks the observations during the primary eclipse 
according to optical ephemerids \citep{gor98}. 

\end{deluxetable}

\begin{deluxetable}{ccccccccccc}
\rotate
\tablewidth{0in}
\tabletypesize{\scriptsize}
    \tablecaption{Best-fit parameters of spectral analysis of PCA and HEXTE
observation of SS~433 in 3-150~keV energy range in the model: {\it wabs*(
bmc+gaussian+gaussian+''bbody'')} 
for observations with numbers 90401-NN-NN-NN.
Parameter errors correspond to 1$\sigma$ confidence level.}
    \renewcommand{\arraystretch}{1.2}
\tablehead{
Model & Parameter & -01-01-00 & -01-01-01 & -01-01-02 & -01-01-03 & -01-02-00 & -01-02-01 & -01-03-00 & -01-03-01 & -01-03-02 }

 \startdata
bmc &                    &        &         &        &         &         &         &             &   &    \\
     & $\Gamma$          & 2.32(2)& 2.30(2) & 2.30(1) & 2.30(2) & 2.30(2) & 2.31(2) & 2.31(2) & 2.29(2) & 2.29(2)\\
     & kT (keV)          & 1.08(2)& 1.15(5) & 1.14(3)& 1.23(5) & 0.9(2)  & 1.18(9) & 1.17(3)  & 1.06(4) & 1.23(5) \\
     & logA$^{\dagger}$   & 0.3(1) & 0.4(2)  & 2.00   &0.3(1) & 1.8(1)    & 0.9(6) & 2.00 & 0.16(2) & 0.4(1) \\
     & N$_{bmc}^{\dagger\dagger}$ & 3.3(6)& 3.4(1) & 3.5(1) & 3.4(1) & 3.6(6) & 3.4(1) & 2.8(1) & 3.2(1) & 3.3(1)     \\
Gaussian1 &               &       &         &              &         &         &         &          &           &  \\
     & E$_{line1}$ (keV)  & 7.06(1)    & 7.23(3)& 7.02(3)  & 7.15(2) & 7.12(2) & 7.07(2) & 7.10(2) & 7.02(3) & 7.04(2) \\
     & $\sigma_{line1}$ (keV)& 0.75(2) & 0.96(8)& 0.87(4)  & 0.80(4) & 0.80(3) & 0.97(4) & 0.90(4)  & 0.70(7) & 0.73(4) \\
     & N$_{line1}^{\dagger\dagger}$ & 4.3(2) & 4.7(6)&4.1(4)& 4.4(8) & 4.3(1)  & 5.3(3)  & 4.2(3)   & 3.7(4)  & 3.8(3)\\
Gaussian2 &               &         &        &          &       &        &        &   &           &  \\
     & E$_{line2}$ (keV)  & 9.08(5) & 9.2(3) & 8.9(2)   &9.3(2) & 9.2(1) & - & - & 8.8(1) & - \\
     & N$_{line2}^{\dagger\dagger}$ &0.30(5)& 0.1(1)&0.1(1)    &0.26(8) & 0.80(3) & - & - & 0.4(1) & -\\
'`bbody'' &              &          &       &              &       &        &         &        &   &  \\
     & T$_{"bbody"}$     & 5.5(1)   & 5.9(2) & 5.9(3)      & 5.3(1)& 5.4(4)  & 5.7(4) & 6.4(6) & 5.4(4) & 5.3(5) \\
     &  (keV)            &          &        &             &       &         &        &        &        &    \\
     & N$_{"bbody"}^{\dagger\dagger}$ & 3.2(1) & 3.6(3)    & 2.8(5) & 4.3(4) & 3.3(5) & 3.3(4) & 1.9(3) & 1.9(3) & 1.1(2)\\
     & EW$_{"bbody"}$    &  1.4(2)  & 1.4(3) &1.4(5)       & 1.4(5) & 1.4(1) & 1.4(2) & 1.4(1) & 1.4(2) & 1.2(1)   \\
     &  (keV)            &          &            &          &       &        &        &        &  &    \\
Flux$^{\dagger\dagger\dagger}$ &    &            &          &       &        &        &        &            & &    \\
     & 3 - 60 keV        & 5.18     & 5.48       & 5.01     & 4.76  & 7.79   & 5.51   & 5.26   & 6.79  & 5.56   \\
     & 13 - 150 keV      & 2.84     & 2.99       & 2.69     & 2.30  & 4.65   & 3.02   & 3.03   & 4.02  & 3.38   \\
      \hline
     & $\chi^2$ (d.o.f.) & 1.24 (69)& 0.98 (69)  & 1.15 (69)& 1.03 (69)&0.97(69) & 1.05 (69) & 1.13 (69) & 1.14 (69)  &  1.08 (69) \\
      \hline
      \enddata
    \label{tab:par_bbody}
$^{\dagger}$when parameter $log(A)>1$, then it  is fixed at  2.0 (see comments in the text), 
$^{\dagger\dagger}$ normalization in units of $10^{-3} (10^{-10}$ ergs/s/cm$^2)$,  
$^{\dagger\dagger\dagger}$ spectral flux 
in units of $10^{-10}$ erg/s/cm$^2$, 
$\sigma_{line2}$ for Gaussian2 component was fixed at  0.01 keV.
\end{deluxetable}


\begin{deluxetable}{cccccccccc}
\tablewidth{0in}
\tabletypesize{\scriptsize}
    \tablecaption{Best-fit parameters of spectral analysis of PCA and HEXTE
observation of SS~433 in 3-150~keV energy range in the model: {\it wabs*(
bmc+gaussian+gaussian+''bbody'')} 
for observations with numbers  91103-NN-NN-NN.
Parameter errors correspond to 1$\sigma$ confidence level.}
    \renewcommand{\arraystretch}{1.2}
\tablehead{
Model & Parameter &-01-02-00 & -01-05-00 & -01-05-01 & -01-06-00 & -01-06-01 & -01-07-00 & -01-08-00&-01-09-01}
 \startdata
bmc &                    &        &         &       &         &         &         &           & \\
     & $\Gamma$          & 2.18(3) & 2.2(1)& 2.20(6) & 2.19(5) & 2.2(1) & 2.21(1) & 2.2(1)  &2.2(1)\\
     & kT (keV)          & 1.04(1)  & 1.29(8)& 1.25(6)& 1.22(7)& 1.23(6) & 1.21(4)  & 1.18(7)&1.11(9) \\
     & logA$^{\dagger}$   &0.2(2) & 0.01(4) & 0.9(1) & 2.0 & 2.0& 2.0 & 2.0 & 2.0 \\
     & N$_{bmc}^{\dagger\dagger}$& 2.21(2) & 2.5(3)& 3.1(4) & 3.2(3) & 3.2(3) & 3.3(2) & 3.0(4) & 3.0(4)\\
Gaussian1 &               &       &         &              &         &         &         &            &\\
     & E$_{line1}$ (keV) &6.70(2) & 7.08(6)    & 7.02(7)& 7.08(5)  & 7.06(3) & 7.03(3) & 7.12(9) & 6.87(6) \\
     & $\sigma_{line1}$ (keV) &0.27(3) & 0.75(8) & 0.75(9)& 0.90(9)  & 0.75(5) & 0.75(6) & 0.98(6) &0.84(9)\\
     & N$_{line1}^{\dagger\dagger}$& 0.6(3) & 2.8(5) & 3.0(5)&3.0(4)& 3.5(3) & 2.8(3)  & 4.1(3) & 3.9(7)\\
Gaussian2 &               &         &       &          &  &     &        &        &    \\
     & E$_{line2}$ (keV) &8.0(2) & - & - & 7.0(1)  &7.01(8) & 7.1(4) & 7.0(1) & -\\
     & N$_{line2}^{\dagger\dagger}$&0.4(1)&-& - &0.05(3) &0.6(3)  & 0.9(2) & 0.5(2)& - \\
'`bbody'' &              &          &       &              &       &        &    &     &         \\
     & T$_{"bbody"}$    & 4.9(2) & 4.9(2)   & 4.9(4) & 5.1(9)      & 5.1(3)& 5.6(5)  & 5.2(4)& 5.5(3) \\
     &  (keV)            &          &        &             &       &         &        &          &\\
     & N$_{"bbody"}^{\dagger\dagger}$ &1.17(7) & 2.1(6) & 1.2(8)    & 0.8(6) & 2.6(5) & 1.7(4) & 1.9(4) & 2.3(6)\\
     & EW$_{"bbody"}$    &  1.2(2)  & 1.3(3) &1.4(5)       & 1.4(5) & 1.4(1) & 1.4(2) & 1.4(1)  & 1.4(2)\\
     &  (keV)            &          &            &          &       &        &        &          &\\
Flux$^{\dagger\dagger\dagger}$ &    &            &          &       &        &        &        &  &           \\
     & 3 - 60 keV       &4.99  & 3.90     & 5.41       & 5.56     & 6.72  & 6.30   & 6.56   & 6.54  \\
     & 13 - 150 keV     &3.17  & 2.13     & 3.34       & 3.44     & 4.01  & 3.88   & 4.29   & 4.34 \\
      \hline
     & $\chi^2$ (d.o.f.) & 1.17(69) & 0.97 (69)& 1.05 (69)  & 1.25 (69)& 1.01 (69)&1.04(69) & 1.03 (69) & 1.05(69) \\
      \hline
      \enddata
    \label{tab:par_bbody}
$^{\dagger}$when parameter $log(A)>1$, it is fixed at  2.0 (see comments in the text), 
$^{\dagger\dagger}$ normalization in units of $10^{-3} (10^{-10}$ ergs/s/cm$^2)$,  
$^{\dagger\dagger\dagger}$ spectral flux 
in units of $10^{-10}$ erg/s/cm$^2$, 
$\sigma_{line2}$ for Gaussian2 component was fixed at  0.01 keV.
\end{deluxetable}



\begin{deluxetable}{ccccccccc}
\tablewidth{0in}
\tabletypesize{\scriptsize}
    \tablecaption{Best-fit parameters of spectral analysis of PCA and HEXTE
observation of SS~433 in 3-150~keV energy range in the model: {\it wabs*(
bmc+gaussian+gaussian+''bbody'')} 
for observations with numbers 91092-NN-NN-NN.
Parameter errors correspond to 1$\sigma$ confidence level.}
    \renewcommand{\arraystretch}{1.2}
\tablehead{
Model & Parameter &-01-01-00 & -01-02-00 & -02-04-00G & -02-05-00 & -02-06-00 & -02-06-01 & -02-08-00}
 \startdata
bmc &                    &        &         &       &         &         &         &                 \\
     & $\Gamma$          & 2.20(3) & 2.20(3)& 2.22(3) & 2.2(3) & 2.22(1) & 2.22(3) & 2.20(1)  \\
     & kT (keV)          & 1.1(1)  & 1.22(3)& 1.20(6)& 1.28(6)& 1.21(7) & 1.3(1)  & 1.35(2) \\
     & logA$^{\dagger}$   &0.6(3) & 0.1(1) & 0.6(4) & 0.5(3) & 1.2(1)    & 0.6(2) & 0.1(1)  \\
     & N$_{bmc}^{\dagger\dagger}$& 2.71(2) & 2.90(5)& 2.7(1) & 2.9(1) & 2.6(2) & 2.67(9) & 2.64(9) \\
Gaussian1 &               &       &         &              &         &         &         &            \\
     & E$_{line1}$ (keV) &6.89(3) & 6.88(1)    & 6.81(2)& 6.83(1)  & 6.90(2) & 6.90(2) & 6.80(6)  \\
     & $\sigma_{line1}$ (keV) &0.75(6) & 0.50(2) & 0.48(5)& 0.46(4)  & 0.51(5) & 0.53(3) & 0.53(2) \\
     & N$_{line1}^{\dagger\dagger}$& 3.7(4) & 2.8(1) & 1.7(1)&2.6(2)& 2.4(2) & 2.5(1)  & 1.6(3) \\
Gaussian2 &               &         &       &          &       &        &        &    \\
     & E$_{line2}$ (keV) &9.1(1) & -     & 8.37(5) & 8.44(1)  &8.4(9) & 8.50(8) & 8.3(2) \\
     & N$_{line2}^{\dagger\dagger}$&0.2(1) & - & 0.40(7) &1.2(4) &2.3(9)  & 0.32(7) & 0.21(2) \\
'`bbody'' &              &          &       &              &       &        &         &         \\
     & T$_{"bbody"}$    & 4.5(2) & 4.5(2)   & 4.4(4) & 4.9(6)      & 5.5(5)& 5.0(4)  & 4.5(4)\\
     &  (keV)            &          &        &             &       &         &        &          \\
     & N$_{"bbody"}^{\dagger\dagger}$ &1.7(7) & 2.2(4) & 1.3(4)    & 2.8(5) & 1.9(6) & 1.7(1) & 1.9(3) \\
     & EW$_{"bbody"}$    &  1.2(2)  & 1.3(3) &1.4(5)       & 1.4(5) & 1.4(1) & 1.4(2) & 1.4(1)  \\
     &  (keV)            &          &            &          &       &        &        &          \\
Flux$^{\dagger\dagger\dagger}$ &    &            &          &       &        &        &        &             \\
     & 3 - 60 keV       &4.36  & 4.07     & 4.42       & 4.43     & 3.61  & 3.98   & 3.35     \\
     & 13 - 150 keV     &2.56  & 2.44     & 2.80       & 2.90     & 1.46  & 1.81   & 1.71    \\
      \hline
     & $\chi^2$ (d.o.f.) & 0.96(69) & 1.07 (69)& 1.39 (69)  & 1.06 (69)& 1.05 (69)&0.96(69) & 1.05 (69)  \\
      \hline
      \enddata
    \label{tab:par_bbody}
$^{\dagger}$when parameter $log(A)>1$, it is fixed at 2.0 (see comments in the text), 
$^{\dagger\dagger}$ normalization in units of $10^{-3} (10^{-10}$ ergs/s/cm$^2)$$\times 10^{-3}$,  
$^{\dagger\dagger\dagger}$ spectral flux 
in units of $10^{-10}$ erg/s/cm$^2$, 
$\sigma_{line2}$ for Gaussian2 component is fixed at  0.01 keV.
\end{deluxetable}


\begin{deluxetable}{cccccccc}
\tablewidth{0in}
\tabletypesize{\scriptsize}
    \tablecaption{BH masses and distances} 
    \renewcommand{\arraystretch}{1.2}
\tablehead{
Source & M$^a_{dyn}$  &  i$_{orb}$ &   i$_{scal}$ & D$^b$ & M$_{scal}$   & D$_{scal}$ & Refs \\
       & (M$_{\odot}$) & (deg)        &  (deg)         & (kpc) & (M$_{\odot}$) & (kpc)     & }
 \startdata
GX~339-4 &   $>$6         &  ...    &  70   & 7.5$\pm$1.6     &  12.3$\pm$1.4     & 5.75$\pm$0.8   & 1, 2 \\
SS~433   &   4.3$\pm$0.8   &  78.7   & 80    & 5.0$\pm$0.5    &  $\gax 2$ & ...    & 3, 4, 5 \\
      \hline
      \enddata
    \label{tab:par_scal}
$^a$ Dynamically determined BH mass, 
$^b$ Source distance found in literature.\\  
References: 
(1) \cite{munos08}, (2) \cite{Hynes04}, (3) \cite{rom87}, (4) \cite{hillwig08}, 
(5) \cite{margon84}. 
\end{deluxetable}

\clearpage
\newpage

\clearpage
\newpage

\clearpage
\newpage

\clearpage
\newpage

\end{document}